\documentclass[onecolumn,amsmath,amssymb]{revtex4}

\usepackage{graphicx}
\usepackage[latin1]{inputenc}
\usepackage{dcolumn}
\usepackage{bm}
\usepackage{epsfig}
\usepackage{amsmath}
\usepackage{color}

\begin{document}


\title{Time reversibility from visibility graphs of non-stationary processes}

\author{Lucas Lacasa and Ryan Flanagan}

\affiliation{School of Mathematical Sciences, Queen Mary University of London, Mile End Road, E14NS, London, UK}%

\date{\today}

\begin{abstract}
Visibility algorithms are a family of methods to map time series into networks, with the aim of describing the structure of time series and their underlying dynamical properties in graph-theoretical terms. Here we explore some properties of both natural and horizontal visibility graphs associated to several non-stationary processes, and we pay particular attention to their capacity to assess time irreversibility. Non-stationary signals are (infinitely) irreversible by definition (independently of whether the process is Markovian or producing entropy at a positive rate), and thus the link between entropy production and time series irreversibility has only been explored in non-equilibrium stationary states. Here we show that the visibility formalism naturally induces a new working definition of time irreversibility, which allows to quantify several degrees of irreversibility for stationary and non-stationary series, yielding finite values that can be used to efficiently assess the presence of memory and off-equilibrium dynamics in non-stationary processes without needs to differentiate or detrend them. We provide rigorous results complemented by extensive numerical simulations on several classes of stochastic processes.
\end{abstract}

\maketitle

\section{Introduction}
A stationary time series ${\cal S}=\{x_1,x_2,\dots,x_n\}$ is called statistically time reversible if the series and its time reversed ${\cal S}^*=\{x_n,x_{n-1},\dots,x_1\}$ are equally likely, i.e. if they have identical joint distributions \cite{lawrance}.
For instance, Gaussian linear processes such as white noise, or conservative chaotic processes such as Hamiltonian chaos are time reversible, and related to processes in thermodynamic equilibrium in statistical physics. Non-linear stochastic processes, or dissipative chaotic processes are generally found to be irreversible \cite{kennel}, and are associated to processes that operate away from equilibrium in a thermodynamic sense. For these cases, recent works relate the amount of entropy that a system is producing while being away from equilibrium to the amount of time irreversibility, computed from the time evolution of adequate physical observables \cite{parrondo}.\\
Traditionally, the study of statistical time irreversibility has only applied to stationary processes \cite{lawrance}. A dynamical process is stationary if its joint distribution does not change under time shift, hence sample time series extracted from the same process at different times have similar statistics, with small deviations only occurring as finite size effects. For these processes, one can then meaningfully estimate properties about the underlying stationary distribution of the process (if this exists) through its estimation for finite series. In particular, one can quantify the amount of time irreversibility in stationary processes via a number of strategies and algorithms proposed in the literature, including simple statistical differences between forward and backward trajectories \cite{parrondo, parrondo2, kennel, costa, cammarota} or more sophisticated methods such as compression \cite{kennel2}. In every case, note that time series need to be symbolized before an irreversibility measure can be computed. Via fluctuation theorems, a remarkable identity between the Kullback-Leibler divergence of the forward and backward statistics of a time series -i.e. the statistics of {\it individual particle trajectories}- and the amount of entropy that the underlying thermodynamic system is producing has been found recently \cite{parrondo, parrondo2}, what has further stimulated the study of time series irreversibility in statistical physics.\\ 
\begin{figure}
\centering
\includegraphics[width=0.5\columnwidth]{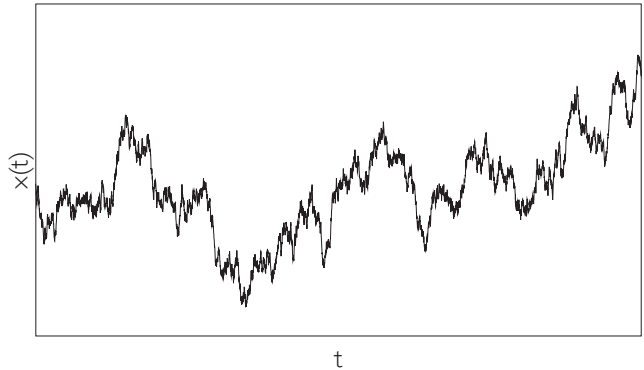}
\caption{Sample time series of an unbiased random walk, as canonical example of a non-stationary process. By definition, the process is (infinitely) time irreversible, although if we remove the Y axis, then it is impossible to know in which direction time is flowing, as both pictures (forward and backward) are equally likely.}
\label{illus}
\end{figure}

\noindent On the other hand, non-stationary processes have underlying joint distributions that change over time, hence no straightforward quantification of the time asymmetry of a process can be extracted from the analysis of finite series. Actually, the precise definition of time series reversibility in time series analysis is the following: a time series ${\cal S}=\{x_1,x_2,\dots,x_n\}$ is called statistically time reversible if the time series ${\cal S}^-=\{x_{-1},x_{-2},\dots,x_{-n}\}$ has the same joint distribution as ${\cal S}$ \cite{weiss}. Of course by definition, non-stationary series are (infinitely) irreversible: the statistical properties of a non-stationary process vary with time, and therefore ${\cal S}$ and ${\cal S}^-$ have different statistics that increase over time without bounds. It is only for stationary processes where the standard definition of time reversibility acquires its full meaning. For this latter case, $\{x_{-1},x_{-2},\dots,x_{-n}\}$ and $\{x_{-1+m},x_{-2+m},\dots,x_{-n+m}\}$ have the same joint distributions $\forall m$, so for the particular choice $m=n+1$, the definition of time reversibility reduces to the equivalence between forward and backward statistics. Hence the popular motto "time reversibility implies stationarity" \cite{lawrance}. Note however, that if we understand the source of irreversibility in close relation to directionality (or, in other words, to underlying sources of memory), then one could argue that there should exist different degrees of irreversibility in non-stationary processes: for instance, a Markovian random walk should arguably be "less irreversible" than a non-Markovian one, even if both are non-stationary. To further illustrate this, in figure \ref{illus} we plot a realization of a 1d random walk $x(t)$ that starts at the origin $x(0)=0$, where we have deliberately removed the vertical axis. While this is a non-stationary process and hence time irreversible, could the reader assert which is the correct direction of time? Wouldn't both the forward and backward processes be equally likely, once the vertical axis is removed? This figure  could indeed both plot $\{x_1\dots x_n\}$ or $\{x_{-1}\dots x_{-n}\}$. Moreover, if $x(t)$ describes the trajectory of a Brownian particle in a system in thermal equilibrium, shouldn't this time series on average have a null entropy production - hence a reversible character?\\

\noindent In this work we show that, by suitably mapping non-stationary time series into a graph-theoretical setting by means of a so called visibility algorithm \cite{PNAS, PRE}, one can actually quantify different kinds of time asymmetries in the underlying dynamics on non-stationary processes, where random walks such as the one presented in figure \ref{illus} are indeed time reversible in the new framework. The family of visibility algorithms were recently introduced as simple mappings between time series and graphs, with the aim of enabling the description and classification of the structure of time series as well as their underlying dynamics in graph-theoretic terms. Among other interesting advantages, these methods do not require the series to be pre-symbolized. In the context of time series irreversibility, a directed version of visibility algorithms was also proposed recently to assess irreversibility in stationary real-valued time series \cite{EPJB, donner}, and has been used extensively \cite{suyal, Xie, donner1, Zou}. Here we extend that former analysis to the realm of non-stationary signals. We investigate the topological properties of so called visibity graphs (VGs) and horizontal visibility graphs (HVGs) associated to several types of non-stationary processes, and pay particular attention to their performance in quantifying several degrees of irreversibility. We take advantage from the fact that the topological properties of these graphs are effectively invariant under time shift for large classes of non-stationary processes, what allows us to introduce the concept of visibility graph stationarity. This in turn allows us to compare to extract meaningful information on the time asymmetry of non-stationary processes.\\ 
 
\noindent The rest of the paper goes as follows: in section II we recall how univariate real-valued time series can be mapped into the family of visibility graphs (natural and horizontal versions), and explain how a directed version of these graphs can be used to estimate statistical time irreversibility of the original time series, without requiring to symbolize the series. We summarize previous findings on canonical stationary processes and prove a lemma that permits us to quantify the degree of irreversibility in non-stationary ones. In section III we focus on random additive processes, and provide some exact results on the properties of visibility graphs associated to simple random walks. We prove that unbiased random walks are indeed time reversible according to new definitions, and that for biased ones, the HVG method can quantify the degree of irreversibility. In section IV we extend these results to random multiplicative processes. We numerically explore the performance of visibility methods in these cases and complement these findings with some analytical and heuristic explanations. In section V we conclude.

\section{Measuring time series irreversibility using visibility graphs}
Here we first introduce the definition of the visibility and horizontal visibility graphs associated to an ordered series of real-valued data. These are inspired in computational geometry \cite{VA} and the intuition underlying the mappings (in particular, the link criteria) shares some similarities with first passage time statistics \cite{redner2}. We also introduce the notions of VG (HVG)-stationarity and VG (HVG)-irreversibility, which we will rely on subsequently.\\

\noindent {\bf Definition (VG). } Let ${\cal S}=\{x_1,\dots,x_n\}$ be a real-valued scalar time (or otherwise ordered) series of $n$ data. A visibility graph (VG) is an undirected graph of $n$ nodes, where each node $i\in [1,n]$ is labelled by the time order of its corresponding datum $x_i$. Hence $x_1$ is mapped into node $i=1$, $x_2$ into node $i=2$, and so on.
Then, two nodes $i$, $j$ (assume $i<j$ without loss of generality) are connected by a link if and only if one can draw a straight line connecting $x_i$ and $x_j$ that does not intersect any intermediate datum $x_k, \ i<k<j$. Equivalently, $i$ and $j$ are connected if the following {\it convexity} criterion is fulfilled:
$$x_k< x_i + \frac{k-i}{j-i}[x_j-x_i],\ \forall k: i<k<j$$
\\
{\bf Definition (HVG). }Let ${\cal S}=\{x_1,\dots,x_n\}$ be a real-valued scalar time (or otherwise ordered) series of $n$ data. A horizontal visibility graph (HVG) is an undirected graph of $n$ nodes, where each node $i\in [1,n]$ is labelled by the time order of its corresponding datum $x_i$. Hence $x_1$ is mapped into node $i=1$, $x_2$ into node $i=2$, and so on.
Then, two nodes $i$, $j$ (assume $i<j$ without loss of generality) are connected by a link if and only if one can draw a {\it horizontal} line connecting $x_i$ and $x_j$ that does not intersect any intermediate datum $x_k, \ i<k<j$. Equivalently, $i$ and $j$ are connected if the following {\it ordering} criterion is fulfilled:\\
$$x_k<\inf(x_i,x_j),\ \forall k: i<k<j$$

\noindent {\bf Definition (VG and HVG stationarity). } A dynamical process $\{X_t\}$ is said to be VG-stationary (HVG-stationary) if and only if the topological properties of the VG (HVG) associated to a sample time series of size $n$ extracted from $\{X_t\}$ are asymptotically (i.e. for large $n$) invariant under time shift (in the statistical sense). In other words, processes for which sample time series $\{x_1,x_2,\dots,x_n\}$ and $\{x_{1+\tau},x_{2+\tau},\dots,x_{n+\tau}\}$ generate (in the limit of large $n$) statistically equivalent VG (HVG) $\forall \tau$ are called VG-stationary (HVG-stationary). In particular, the degree distributions of VG (HVG) associated to $\{x_1,x_2,\dots,x_n\}$ and $\{x_{1+\tau},x_{2+\tau},\dots,x_{n+\tau}\}$ are asymptotically (for large $n$) identical for VG (HVG) stationary processes.\\

\noindent {\bf Lemma 1.} Let $\{X_t\}$ be a non-stationary process, and consider two time series samples of $n$ data extracted from $\{X_t\}$: $\{x_1,x_2,\dots,x_n\}$ and $\{x_{1+\tau},x_{2+\tau},\dots,x_{n+\tau}\}$ for some $\tau \in \mathbb{Z}$. If $\forall \tau  \  \exists c \in \mathbb{R}$ such that $\{x_1,x_2,\dots,x_n\}$ and $\{x_{1+\tau}+c,x_{2+\tau}+c,\dots,x_{n+\tau}+c\}$ are statistically equivalent time series (i.e. have the same joint distributions), then the process $\{X_t\}$ is VG (HVG) stationary. \\

\noindent {\it Proof:} Both VG and HVG are invariant under vertical rescaling of the time series \cite{PNAS}, that is to say, the series ${\cal S}=\{x_1,x_2,\dots,x_n\}$ and ${\cal S}'=\{x_1+c,x_2+c,\dots,x_n+c\}$ generate the same VG and HVG $\forall c \in \mathbb{R}$. Thus $\{x_{1+\tau}+c,x_{2+\tau}+c,\dots,x_{n+\tau}+c\}$ and $\{x_{1+\tau},x_{2+\tau},\dots,x_{n+\tau}\}$
also generate the same VG and HVG, $\forall c\in \mathbb{R}$. Choose $c$ such that $\{x_1,x_2,\dots,x_n\}$ and $\{x_{1+\tau}+c,x_{2+\tau}+c,\dots,x_{n+\tau}+c\}$ are statistically equivalent (they have identical asymptotic joint distributions). Note that this usually is fulfilled in additive processes for $c=x_1-x_{1+\tau}$, but should be set for each process independently.
Then, $\{x_{1+\tau}+c,x_{2+\tau}+c,\dots,x_{n+\tau}+c\}$ and $\{x_1,x_2,\dots,x_n\}$ will also generate statistically equivalent VG and HVG, hence by definition the process is VG and HVG stationary $\square$.\\

\noindent Both the VG and HVG have been fruitfully applied in recent years to describe and classify different types of time series and dynamics. For instance, VG have been shown to be a viable method to quantify the Hurst exponent of fractional Brownian motion (inherently non-stationary signals), as 
a linear relation was found between the Hurst exponent $H$ of a time series and the exponent $\gamma$ of the power law degree distribution of the associated VG, $\gamma=3-2H$ \cite{EPL}. The HVG has been used in turn to describe chaotic and correlated stochastic processes \cite{toral}, or to provide a graph-theoretical description of canonical routes to chaos \cite{chaos1, chaos2, chaos3}, and it has been shown that HVGs are analytically tractable for several classes of Markovian dynamics \cite{nonlinearity}.
Both VG and HVG are connected planar graphs by construction, which have a hamiltonian path described by the path $1-2-\dots-n$. HVG are indeed outerplanar graphs, and again by construction, one can easily prove that the HVG of $\cal S$ is a subgraph of VG. As both VG and HVG have a natural order induced by the time arrow (or equivalently, by the order of the associated series $\cal S$), it is natural to define the degree sequence of a VG or a HVG as $\{k(t)\}_{t=1}^n$, where $k(t)$ is the degree of node $i=t$. 
\\
Note that previous definitions generate undirected graphs. However, these can be made {\it directed} by again assigning to the links a time arrow. Accordingly, a link between $i$ and $j$ (where time ordering yields $i<j$), generates an {\it outgoing} link for $i$ and an {\it ingoing} link for $j$. The degree sequence thus splits into an ingoing degree sequence $\{k_{in}(t)\}_{t=1}^n$, where $k_{in}(t)$ is the ingoing degree of node $i=t$, and an outgoing degree sequence. An important property at this point is that the ingoing and outgoing degree sequences are interchangeable under time series reversal. That is to say, if we define the time reversed series ${\cal S}^*=\{x_{n+1-t}\}_{t=1}^n$, then we have the following identities
\begin{equation}
\{k_{in}(t)\}[{\cal S}]=\{k_{out}(t)\}[{\cal S}^*]; \ \{k_{out}(t)\}[{\cal S}]=\{k_{in}(t)\}[{\cal S}^*]
\label{degseqrev}
\end{equation}
Now, one can define, from the ingoing and outgoing degree sequences, 
an ingoing degree distribution $P(k_{in})\equiv P_{in}(k)$ and an outgoing degree distribution $P(k_{out})\equiv P_{out}(k)$, and property (\ref{degseqrev}) is inherited in the distributions, such that
\begin{equation}
P_{in}(k)[{\cal S}]=P_{out}(k)[{\cal S}^*]; \ P_{out}(k)[{\cal S}]=P_{in}(k)[{\cal S}^*]
\label{degdisrev}
\end{equation}

\noindent {\bf Definition (VG and HVG reversibility)} In this work, a time series ${\cal S}=\{x(t)\}_{t=1}^n$ is said to be (order $p$) VG-reversible (HVG-reversible) if and only if, for large $n$, the order $p$ block {\it in} and {\it out} degree distribution estimates of the VG (HVG) associated to ${\cal S}$ are asymptotically identical, i.e.
$$P_{in}(k_1k_2\dots k_p)= P_{out}(k_1k_2\dots k_p).$$
Note that according to property (\ref{degdisrev}), the previous formula implies that, under the VG/HVG setting, the statistics of the degree sequences are statistically invariant under time reversal.  Other topological properties of VG/HVG could be used to quantify time asymmetries, as has been reported recently \cite{donner}. For finite series, we will assess how close the system is to reversibility by quantifying the distance (in distributional sense) between $P_{in}$ and $P_{out}$. 
While several possible measures can be used, here we focus in the so called Kullback-Leibler divergence between the {\it in} and {\it out} distributions, previously proposed in \cite{EPJB}:
$$D_{kld}(\text{in}||\text{out})=\sum_kP_{in}(k)\log \bigg(\frac{P_{in}(k)}{P_{out}(k)}\bigg)$$\\
$D_{kld}(in||out)$ is a semi-distance which is null if and only if $P_{in}(k)=P_{out}(k)$, and is positive otherwise. We deal with the situation where $P_{in}(k)\neq 0, \ P_{out}(k)=0$ by introducing a small bias that removes finite-size divergences, as suggested recently \cite{parrondo2}. We then redefine VG/HVG-reversibility as 
$$\lim_{n\to\infty}D_{kld}(\text{in}||\text{out})=0$$
Truly irreversible processes will have positive irreversibility values even in the limit of large $n$: we will call these processes VG/HVG-irreversible. For VG/HVG-reversible processes, $D_{kld}(in||out)$ will have a positive finite value for finite size series that vanishes as size increases as $n^{-\delta}$ (where $\delta$ will be different for VG and HVG). As the convergence is relatively slow, the finite-size results will also be helpful to compare and classify the degrees of reversibility of finite series across different processes, something relevant in practice.\\ 
We have checked that all the results we found in this work are qualitatively equivalent under alternative distance measures between distributions, such as the Manhattan ($L_1$) distance $D_{L1}=\sum_k |P_{in}(k)-P_{out}(k)|$, although in this latter case, convergence to zero for reversible cases is typically slower (results not shown). We also chose the Kullback-Leibler divergence one as it has some physical meaning: for stationary series, $D_{kld}(in||out)$ provides a lower bound \cite{EPJB} to the thermodynamic entropy that a non-equilibrium steady state described by a state variable $x(t)$ is producing along its time evolution. 
Also, as {\it in} degrees account for past information while {\it out} degrees account for future information (or past information in the time reversed case), then $D_{kld}(\text{in}||\text{out})$ is formally akin to $D_{kld}(\text{forward}||\text{backwards})$ in graph space, whereas or $D_{kld}(\text{out}||\text{in})$ is the formal analog to $D_{kld}(\text{backwards}||\text{forward})$.
This measure was used to assess HVG-reversibility in the context of stationary processes and non-equilibrium steady states. Here we further extend that analysis to investigate both HVG and VG reversibility for several classes of dynamics. In what follows, we drop the specification and in the text we refer to $D_{kld}(\text{in}||\text{out})\equiv D$.\\

\noindent In this work we will mainly look at $p=1$, so for readability we drop this specification from now on. Some important remarks are in order. First, note that there is no direct equivalence between order $p$ VG (HVG) reversibility in stationary processes and  order $p$ reversibility in the time series, expressed as $P(x_1,\dots,x_p)=P(x_p,\dots,x_1)$. As a matter of fact, the degree of each node in a VG(HVG) graph inherits information from the whole time series, hence it is a global measure. Nonetheless, as we only look at order $p=1$, we can't rule out the possibility that certain processes appear to be VG/HVG reversible at order $p=1$ but are found to be irreversible at higher orders, as happens for time series produced out of equilibrium where the net current is balanced to zero via stalling forces \cite{EPJB}. So whereas in this work we are dropping the 'order $p$' for readability, the reader should recall that we are working at order $p=1$ in the VG/HVG setting.\\
Second, it is important to highlight that standard methods that aim to quantify time series reversibility usually address the statistical differences of time series directly. As already stated, the original definition of time reversibility precludes the possibility of quantifying irreversibility in non-stationary signals --${\cal S}=\{x_1,x_2,\dots,x_n\}$ and ${\cal S^-}=\{x_{-1},x_{-2},\dots,x_{-n}\}$ have statistical differences that grow with $n$ for non-stationary processes.\\
Third, in order to assess irreversibility directly in real-valued data, one is unavoidably required to {\it symbolize} the series in advance: one needs to pre-define an alphabet (with an arbitrary number of symbols) and generate a time series partition to map each datum into a symbol. Both the alphabet and the partition have to be defined ad hoc, and results often depend on these free parameters, what inevitably generates ambiguities in finite size. Furthermore, in the non-stationary realm, symbolization is clearly ill-defined as the phase space itself grows with the series size.\\
Here, we take advantage of the properties of the visibility algorithms, and apply the irreversibility measures directly on the degree sequences $\{k(t)\}_{t=1}^T$, where $k(i)$ is the degree of node $i$. This sequence is discrete by construction, so there is no need to perform any ad hoc symbolization.

\subsection {Preamble on stationary systems: white noise versus fully developed chaos.}
\begin{figure}
\centering
\includegraphics[width=0.48\columnwidth]{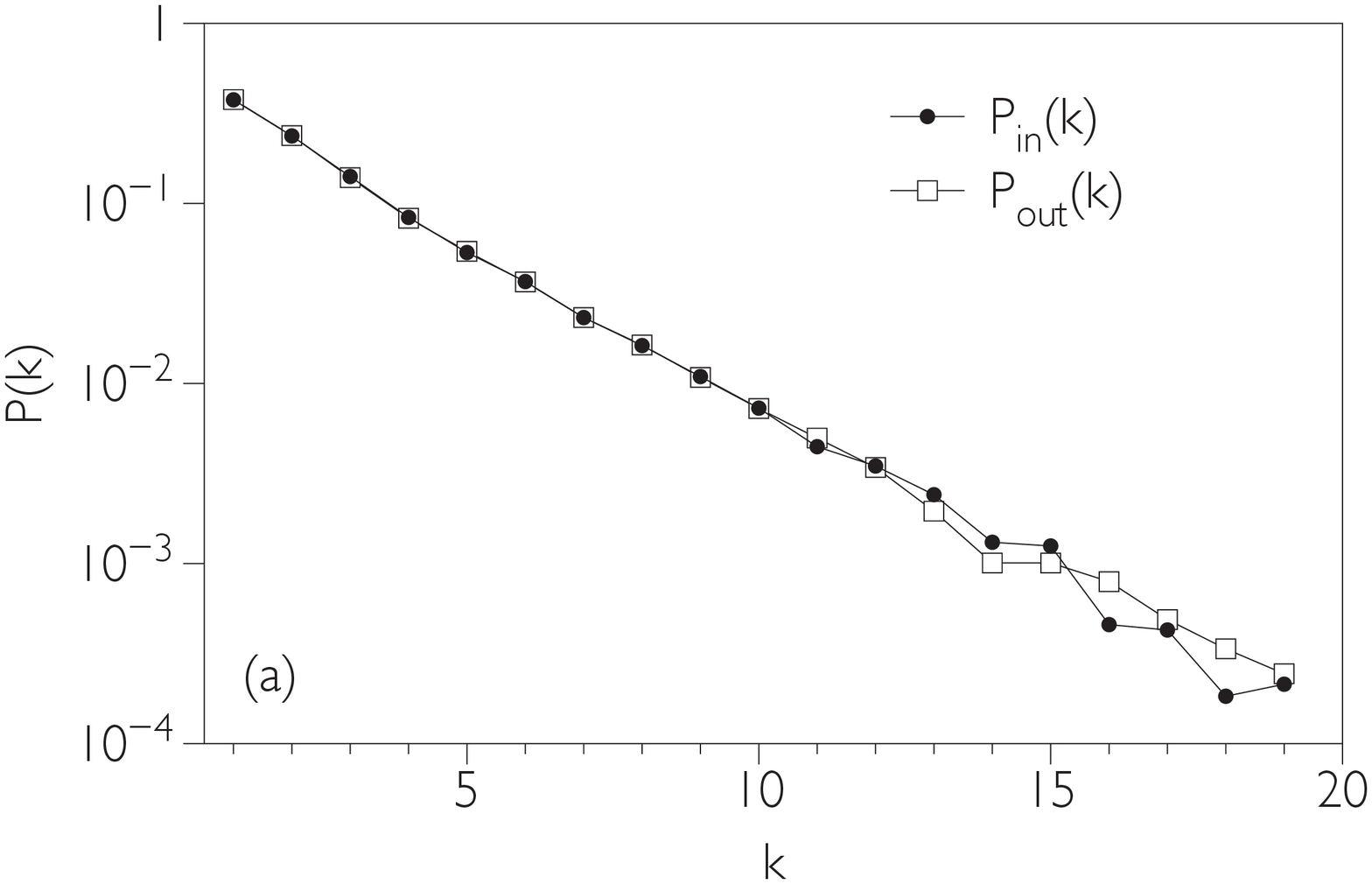}
\includegraphics[width=0.48\columnwidth]{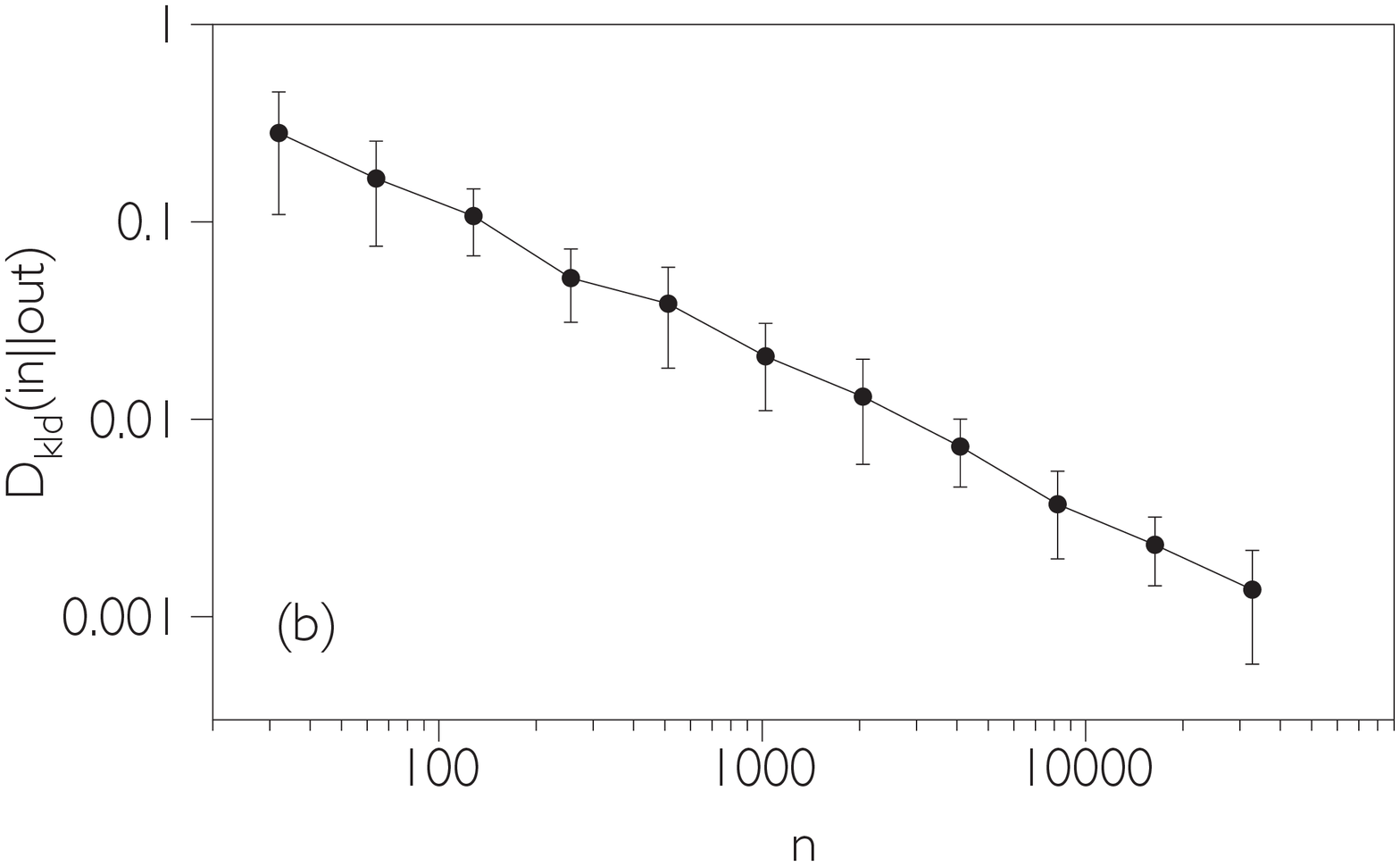}
\caption{{\bf White noise - VG. }(a) Semi-log plot of the in and out degree distributions of the natural visibility graph associated to a time series of $2^{15}$ i.i.d. uniformly random uncorrelated variables $\sim U[0,1]$. Both distributions are identical up to finite-size effects fluctuations, suggesting that the underlying process is VG-reversible. (b) Log-log plot of the irreversibility measure $D_{kld}(in||out)$ as a function of the series size $n$ (each dot is an average over 10 realizations). This measure vanishes asymptotically as $1/n$, showing that finite irreversibility values for finite size are due to statistical fluctuations that vanish asymptotically.}
\label{whitenoise}
\end{figure}

As an illustration, let us begin by considering two paradigmatic stationary processes. The first one is white noise, a stationary and statistically time reversible uncorrelated stochastic process. Consider a sequence of i.i.d. uncorrelated random variables (i.e. $\langle \xi(t)\xi(t+\tau)\rangle=\delta(\tau)$) extracted from some probability distribution $p(x)$ with some compact real support as a realization of the white noise process. For this process, a theorem \cite{EPJB} guarantees that asymptotically $P_{in}(k)=P_{out}(k)=2^{-k}$ and as a consequence, the process is HVG-reversible $\forall p(x)$. As we lack equivalent theorems for VGs, we have run numerical simulations. In figure \ref{whitenoise} we plot, in semi-log, $P_{in}(k)$ and $P_{out}(k)$ for a VG associated to a sample of $2^{15}$ i.i.d. uniform random variables $\sim U[0,1]$. In panel (b) of the same figure we plot the irreversibility estimate $D$ for increasing system size (each dot is an average over 10 realizations). We conclude that white noise is both HVG and VG-reversible showing that this process is indeed VG-reversible, in good agreement with previous theory.\\

\begin{figure}
\centering
\includegraphics[width=0.48\columnwidth]{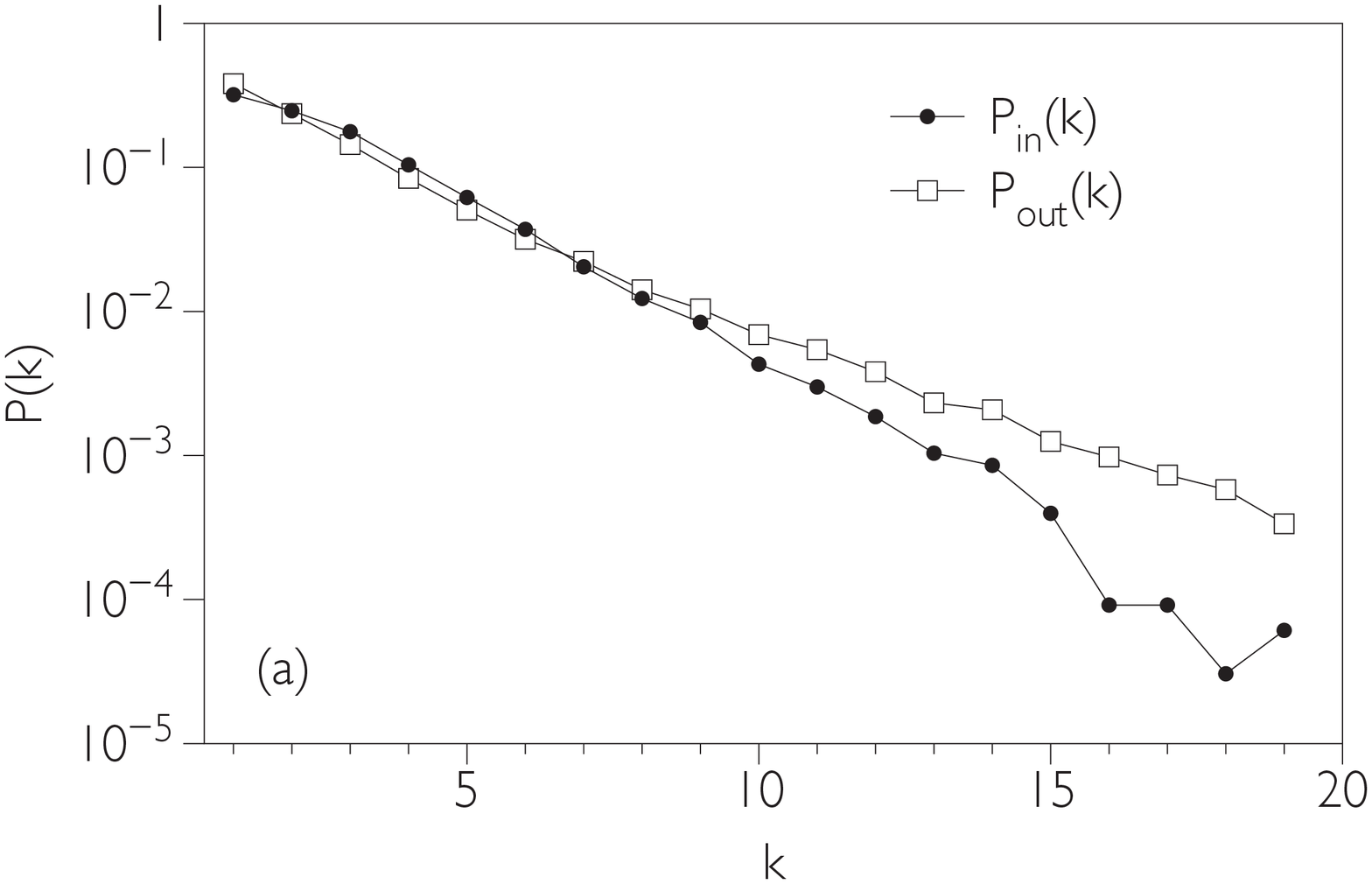}
\includegraphics[width=0.48\columnwidth]{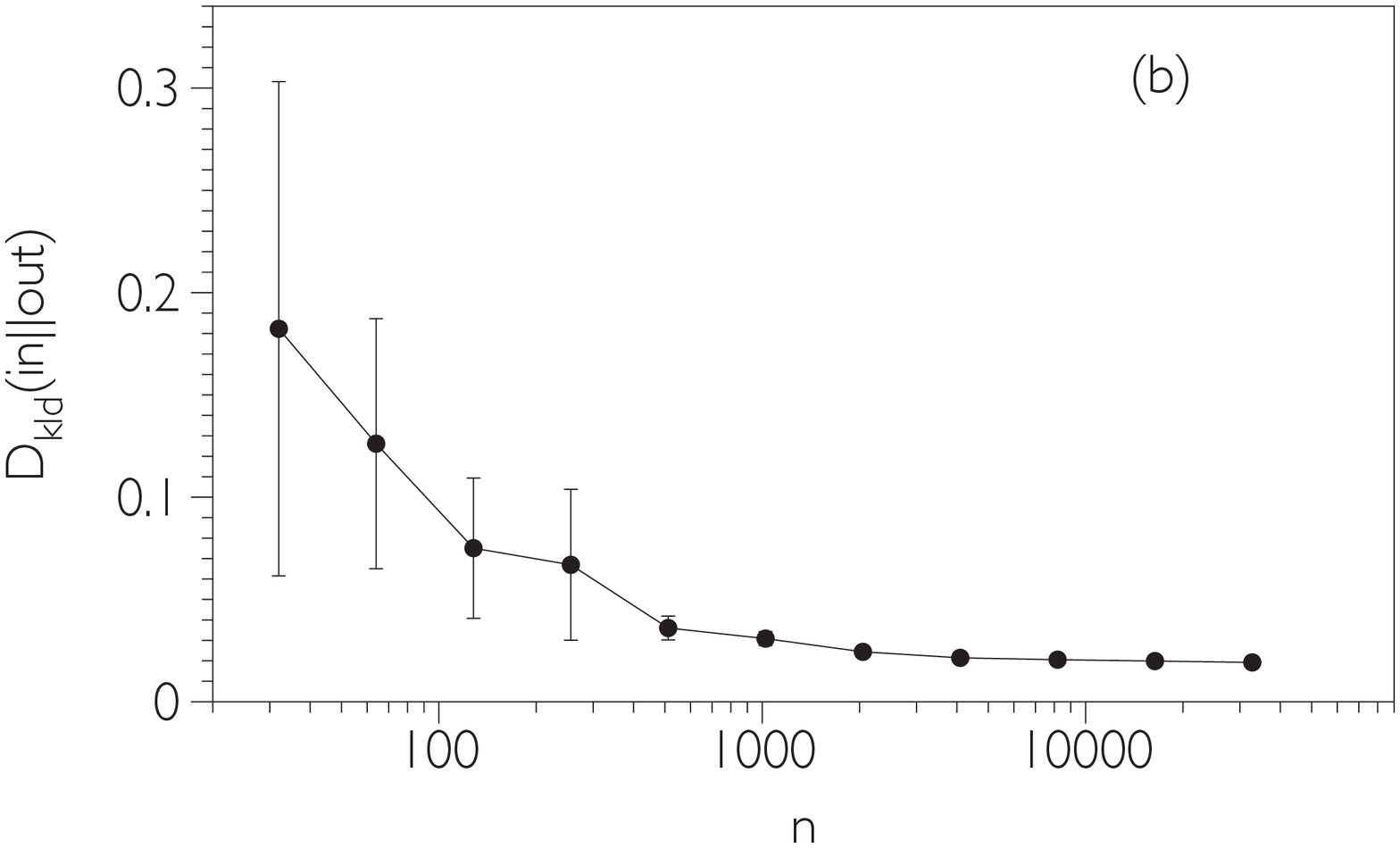}
\caption{{\bf Dissipative chaos - VG. }(a) Semi-log plot of the in and out degree distributions of the natural visibility graph associated to a time series of $2^{15}$ data generated from a fully chaotic logistic map $x(t+1)=4x(t)(1-x(t))$. Distributions are clearly different, suggesting that the underlying process, although stationary, is VG-irreversible. (b) Irreversibility measure $D_{kld}(in||out)$ as a function of the series size $n$ (each dot is an average over 10 realizations). This measure converges to a finite value with series size, confirming that the process yields a positive irreversibility measure.}
\label{logistic}
\end{figure}

\noindent For comparison, we also consider a fully chaotic logistic map $x_{t+1}=4x_t(1-x_t)$, where $x \in [0,1]$, a paradigmatic deterministic stationary process which is nonetheless time irreversible. HVG-irreversibility of the fully chaotic logistic map was shown in \cite{EPJB}, where it was found that $P_{in}(k)$ and $P_{out}(k)$ were asymptotically different distributions. We can summarize this by computing $P_{in}(1)$ and $P_{out}(1)$, and showing that they are strictly different. We first rely on the fact that this map is Markovian, hence\\

\begin{eqnarray}
P_{out}(k=1) =\int_0^1 dx_{t} \int_{x_t}^1 dx_{t+1} f(x_{t})f(x_{t+1}|x_t), \nonumber\\
P_{in}(k=1)=\int_0^1 dx_t \int_{x_t}^1 dx_{t-1} f(x_{t-1}) f(x_t|x_{t-1}). \nonumber
\end{eqnarray}
where $f(x)$ is the invariant probability measure that
characterizes the long-term fraction of time spent by the system in
the various regions of the attractor. In the case of the (fully chaotic) logistic
map the attractor is the whole interval $[0,1]$ and the
invariant measure is
\begin{equation}
f(x)=\frac{1}{\pi \sqrt{x(1-x)}}. \label{rho}
\end{equation}
Now, for a deterministic system, the transition probability is
simply
\begin{equation}
f(x_{t+1}|x_t)=\delta(x_{t+1}-F(x_t)), \nonumber \label{rho2}
\end{equation}
where $\delta(x)$ is the Dirac delta distribution and $F(x)=4x(1-x)$.
Notice that, using the properties of the Dirac delta distribution,
$\int_{x_t}^1 \delta(x_{t+1}-F(x_t))dx_{t+1}$ is equal to one if and only if
$F(x_t)\in[x_t,1]$, what happens for $0<x_t<3/4$, and it is zero
otherwise. Therefore the only effect of this integral is to
restrict the integration range  of $x_t$ to be $[0,3/4]$:
\begin{equation}
P_{out}(k=1)= \int_0^{3/4}dx_t f(x_t)=2/3. \label{poutlog} \nonumber
\end{equation}
Similarly,
$$P_{in}(k=1)=\int_{3/4}^1 f(x_t)dx_t=1/3.$$
 We conclude that $P_{out}(1)\neq P_{in}(1)$ for the fully chaotic
logistic map. Since $D$ is semi-positive definite and null if and only if the two distributions are identical, then $D$ is strictly positive for this process, i.e. it is HVG-irreversible.\\
As we don't have equivalent theory for VGs, we have again run numerical simulations for this case, which are plotted in figure \ref{logistic}. Once again, the {\it in} and ${\it out}$ distributions are clearly different and their Kullback-Leibler divergence converges to a finite, positive value as the series size increases, also suggesting VG-irreversibility.\\

\noindent In what follows we extend previous studies on stationary signals to the realm of non-stationary time series. 

\section{Additive random walks}
\subsection{Simple random walks}
Let us start by considering a simple one dimensional random walk, described by
\begin{equation}
x(t+1)=x(t)+\xi, \ \xi \in \{-1,1\}, 
\label{RWdic}
\end{equation}
i.e. the step distribution is the Rademacher-1/2 distribution.
Without loss of generality, if we generate a time series of $n$ data $\{x_1,\dots,x_n\}$ which deterministically starts in the origin (for which $\mathbb{E}(x_1)=\sigma^2(x_1)=0$), as the process is unbiased, we have $\mathbb{E}(x_n)=0 \ \forall n$, but, in virtue of the central limit theorem, the variance fulfills $\sigma^2(x_n)\sim n$, so the process is non-stationary. In this subsection we derive exact results on the in/out degree distributions for this simple process.\\

\noindent{\bf Theorem 1.} The {\it in} and {\it out} degree distributions of the HVG associated to a bi-infinite series generated by a 1d simple random walk are 
\begin{equation}
P_{in}(k)=P_{out}(k)=
\left\{
\begin{array}{rcl}
     1/2 & k=1,2\\
     0 & \textrm{otherwise}
\end{array}
\right.
\label{theo}
\end{equation}

\noindent {\bf Proof.} First, notice that we don't necessarily need to compute $P_{out}(k)$ and $P_{in}(k)$ separately. We use property \ref{degdisrev} and focus on both $P_{out}$ as applied to the time series and its time reverse, that we label $P_{out}$ and $P^*_{out}$ respectively.\\
\begin{itemize}
\item $k=0$: by construction there is exactly one node with $k_{out}=0$ (the final node) and only one node with $k_{in}=0$ (the initial node), so $P_{out}(0)=P_{out}^*(0)=1/N$, where $N$ is the series size. Hence for bi-infinite series, $P_{out}(0)=P_{out}^*(0)=0$.
\item $k=1$: $P_{out}(1)=\text{prob}(x_{t+1}\geq x_t)=1/2$, $P_{out}^*(1)=\text{prob}(x_{t}\geq x_{t+1})=1/2$. 
\item $k>2$: let us prove by contradiction that $P_{out}(k>2)=P_{out}^*(k>2)=0$. In order for $P_{out}(k>2)>0$, there should be at least an ordering of data that allows that a node chosen at random has degree $k_{out}>2$. Let us assume that the node associated to datum $x_0$ is that node, which {\it at least} has {\it out} visibility of the node associated to $x_1$ (by construction), $x_p$ (for some $1<p<q$) and $x_q$ (for some $q>2$). The geometrical restrictions on the data that follow from the horizontal visibility criterion are $\{x_0>x_1; \ x_0>x_p>x_1, \  x_q>x_p\}$. The first restriction  yields $x_1=x_0-1$ according to equation \ref{RWdic}. On the second condition we have $x_p>x_1$ that implies $x_p\geq x_0$. But this contradicts the first inequality of the second restriction, $x_0>x_p$. Hence $P_{out}(k>2)=0$.
A similar geometrical argument yields $P_{out}^*(k>2)=0$.
\item Normalization of the probability yields $P_{out}(2)=P_{out}^*(2)=1/2$, what concludes the proof.$\ \square$  
\end{itemize}

\noindent{\bf Explicit computation of $P_{out}(2)$: Dyck words.}\\
In the previous proof, we didn't need to compute explicitly $P_{out}(2)$. As a curiosity, here we show that this is indeed possible using simple enumerative combinatoric arguments. We start by using the diagrammatic approach proposed in \cite{nonlinearity}, which divides the computation of each degree probability into an infinite sum of corrections of order $\alpha$, $P_{out}(2)=\sum_{\alpha=0}^\infty P_{out}^{(\alpha)}(2)$, where $\alpha$ is the number of hidden variables (hidden data) in a given configuration. That is to say, $P_{out}^{(\alpha)}(2)$ gathers the contribution given by all the diagrams for which we find $k_{out}=2$, that include a total of $\alpha$ hidden variables (hidden data with no visibility). For instance, for $k_{out}=2$ there is exactly one path (diagram) at order $\alpha=0$, that can be labeled as $\{BT\}$, where $B$ stands for a movement downhill ($\xi=-1$) and $T$ stands for a single movement uphill ($\xi=+1$). This represents the diagram $\{x_0,x_1,x_2\}$ where $x_1=x_0-1,\ x_2=x_0$, and its associated probability is directly $2^{-2}$. There are no contributing paths at order $\alpha=1$ (actually all odd values of $\alpha$ are forbidden by construction), whereas there is exactly one path at order $\alpha=2$, labeled as $\{BBTT\}$, that contributes with a probability $2^{-4}$.
Actually, any path should start and end by $\{ B| \cdots |T \}$. The number of hidden variables $\alpha$ is represented here as the number of extra letters to be located. While there are a total of $2^{\alpha}$ possible paths that start with a downhill movement and end with an uphill movement (with equal weight $2^{-(\alpha+2)}$), not all of them are allowed in the sense of generating a valid path for $k_{out}=2$ - only strictly negative closed walks of length $\alpha +2$ are allowed at order $\alpha$.\\ First, $k_{out}=2$ requires that the initial and final node have associated data of identical height. Since the initial movement is downhill ($B$) and the final one is uphill ($T$), the hidden variables should contribute with a null vertical movement, so half of them have to be involved in a downhill movement, and half of them in an uphill one. This reduces the number of paths from $2^\alpha$ to ${\alpha \choose \alpha/2}$. Furthermore,
only those paths that always remain under $x_0$ until reaching the end datum will actually be paths of order $\alpha$ (if they cross the $x_0$ level at prior stages they are considered corrections of lower order). 
Interestingly, the number of allowed paths can then be seen as the number of words of length $\alpha$ having $\alpha/2$ B's and $\alpha/2$ T's, such that no initial segment of the word has more T's than B's. These paths are sometimes called {\it Dyck words} in enumerative combinatorics. The number of Dyck words of size $\alpha$ is ${\cal C}_{\alpha/2}$, where ${\cal C}_n$ is the Catalan number
$${\cal C}_n=\frac{1}{n+1}{2n \choose n}$$
Hence $P_{out}(2)$ takes the form \begin{eqnarray}
P_{out}(2)&=&\sum_{\alpha=0,\text{even}}^{\infty} {\cal C}_{\alpha/2}\bigg(\frac{1}{2}\bigg)^{\alpha+2}=\frac{1}{4}\sum_{\gamma=0}^{\infty} \frac{1}{\gamma+1} {2\gamma \choose \gamma} \bigg(\frac{1}{4}\bigg)^{\gamma}
\label{p2final}
\end{eqnarray}
where we have used the change of variable $\gamma=\alpha/2$. Leaving the pre factor $1/4$ aside, equation (\ref{p2final}) is the generating function of the Catalan numbers evaluated at $z=1/4$. The generating function sums up to $[1-\sqrt{1-4z}]/2z$, thus
$$P_{out}(2)=1/2,$$
in good agreement with previous theorem.\\

By virtue of lemma 1, the process described in equation (\ref{RWdic}) is VG and HVG-stationary (choosing $c$ such that every sample time series starts, say, a the origin, makes them statistically indistinguishable). Accordingly, one is entitled to explore the time asymmetries taking place in the graph space. According to theorem 1, as both the in and out degree distributions are equivalent for the HVG, the process is indeed HVG-reversible. In what follows we explore a generalization of this process and the performance of both HVG and VG.

\subsection{Unbiased additive random walks}

Let us generalize the previous simple random walk by considering an unbiased additive random walk
\begin{equation}
x(t+1)=x(t)+\xi, \  \  \langle \xi\rangle=0
\label{RWmodel}
\end{equation}
where $\xi$ are i.i.d. random variables extracted from some (arbitrary) symmetric distribution. This process is, for instance, a (1d) discrete model of a Brownian particle evolving in an infinitely large system which is in thermodynamic equilibrium, a system which on average is not producing entropy. From a time series perspective, it is however a non-stationary process, i.e. time irreversible. 
The following theorem uses the VG and HVG method to somehow reconcile both aspects.\\ 

\noindent {\bf Theorem 2.} A bi-infinite time series generated from the unbiased random walk model defined in equation \ref{RWmodel} is both VG and HVG reversible.\\

\noindent {\it Proof:}
The first step is to prove that the process described in equation \ref{RWmodel} is both VG and HVG stationary. Choose $c=x_1-x_{1+\tau}$ in lemma 1, for which $$\{x_{1+\tau}+c,x_{2+\tau}+c,\dots,x_{n+\tau}+c\}=\{x_1,x_1+(x_{2+\tau}-x_{1+\tau}),\dots,x_1+ (x_{n+\tau}-x_{1+\tau})\}=\{x_1,x_1+\xi,\dots,x_1+\sum_{i=1}^{n-1}\xi\}.$$ This last series is equivalent by definition to $\{x_1,x_2,\dots,x_n\}$, therefore the process is VG and HVG stationary, concluding the first part of the proof.\\
Accordingly, reversibility reduces to investigate whether the VG/HVG of $\{x_1,x_2,\dots,x_n\}$ and $\{x_n, x_{n-1},\dots,x_1\}$ are statistically identical. To address this, we recall that visibility algorithms (both VG and HVG) are invariant under vertical rescaling \cite{PNAS, PRE}. This means that two time series $\{x_1,x_2,\dots,x_n\}$ and $\{x_1+c,x_2+c,\dots,x_n+c\}$ yield the same VG (and the same HVG) $\forall c \in \mathbb{R}$. In particular, the (vertically shifted) reversed time series $\{x_n + c,x_{n-1} +c\dots,x_1+c\}$ and the reverse time series $\{x_n, x_{n-1}\dots,x_1\}$ also yield the same VG and HVG $\forall c \in \mathbb{R}$. Our strategy then consists in proving that there exists a value $c$ for which $\{x_n + c,x_{n-1} +c\dots,x_1+c\}$ and $\{x_1,x_2,\dots,x_n\}$ are statistically identical. Choose $c=x_1-x_n$, for which $\{x_n + c,x_{n-1} +c\dots,x_1+c\}=\{x_1,x_{1}-(x_n-x_{n-1}),x_1-(\xi + x_{n-1}-x_{n-2}),\dots \}=\{x_1,x_{1} -\xi, x_1- \sum_{i=1}^2 \xi, \dots,x_1-\sum_{i=1}^{n-1}\xi\}$. Note that in the last series, since $\xi$ has a symmetrical distribution for the process under study, then it is invariant under the transformation $\xi\to -\xi$. So $\{x_1,x_{1} -\xi, x_1- \sum_{i=1}^2 \xi, \dots,x_1-\sum_{i=1}^{n-1}\xi \}$ and $\{x_1,x_{1} +\xi, x_1+ \sum_{i=1}^2 \xi, \dots,x_1+\sum_{i=1}^{n-1}\xi \}$ are indeed statistically equivalent. But this latter series is equivalent by definition to $\{x_1,x_2,\dots,x_n\}$, thus concluding that the process described in equation \ref{RWmodel} is both VG and HVG reversible. $\square$ \\

\begin{figure}
\centering
\includegraphics[width=0.48\columnwidth]{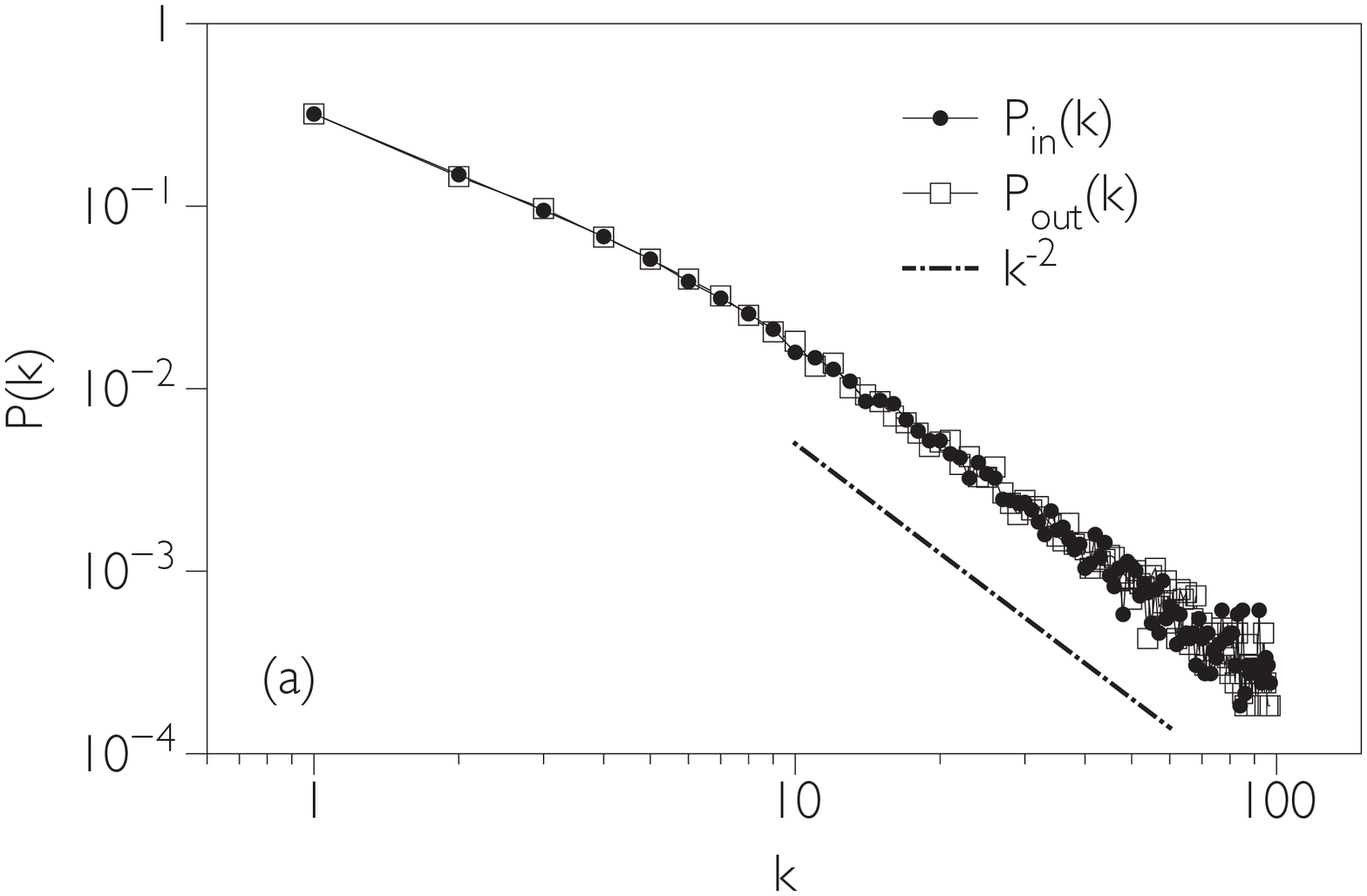}
\includegraphics[width=0.48\columnwidth]{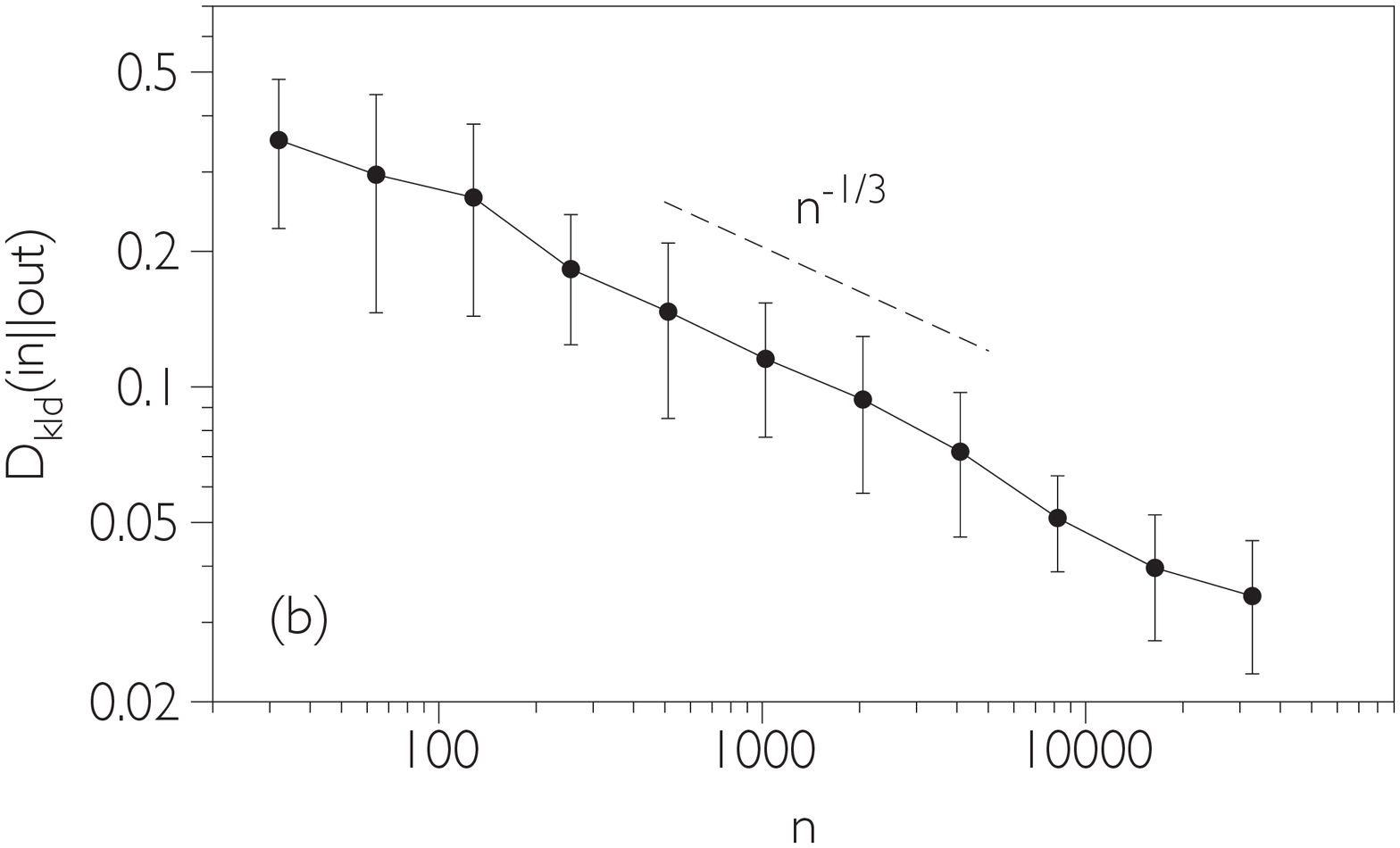}
\caption{{\bf Non-stationary unbiased (memoryless) additive random walk - VG. }(a) Log-log plot of the in and out degree distributions of the natural visibility graph associated to an unbiased random walk of $2^{17}$ steps generated from $x(t+1)=x(t)+\xi$, where $\xi \sim U[-0.5,0.5]$. Both distributions are identical up to finite-size effects fluctuations, suggesting that the underlying process is VG-reversible. The distributions follow a power law tail $k^{-2}$, something that can be heuristically justified according to scaling laws (see the text).
(b) Log-log plot of the irreversibility measure $D_{kld}(in||out)$ as a function of the series size $n$ (each dot is an average over 10 realizations). This measure vanishes asymptotically as $n^{-1/3}$, suggesting that, albeit being a non-stationary process, it is VG-reversible.}
\label{RW}
\end{figure}

\begin{figure}
\centering
\includegraphics[width=0.48\columnwidth]{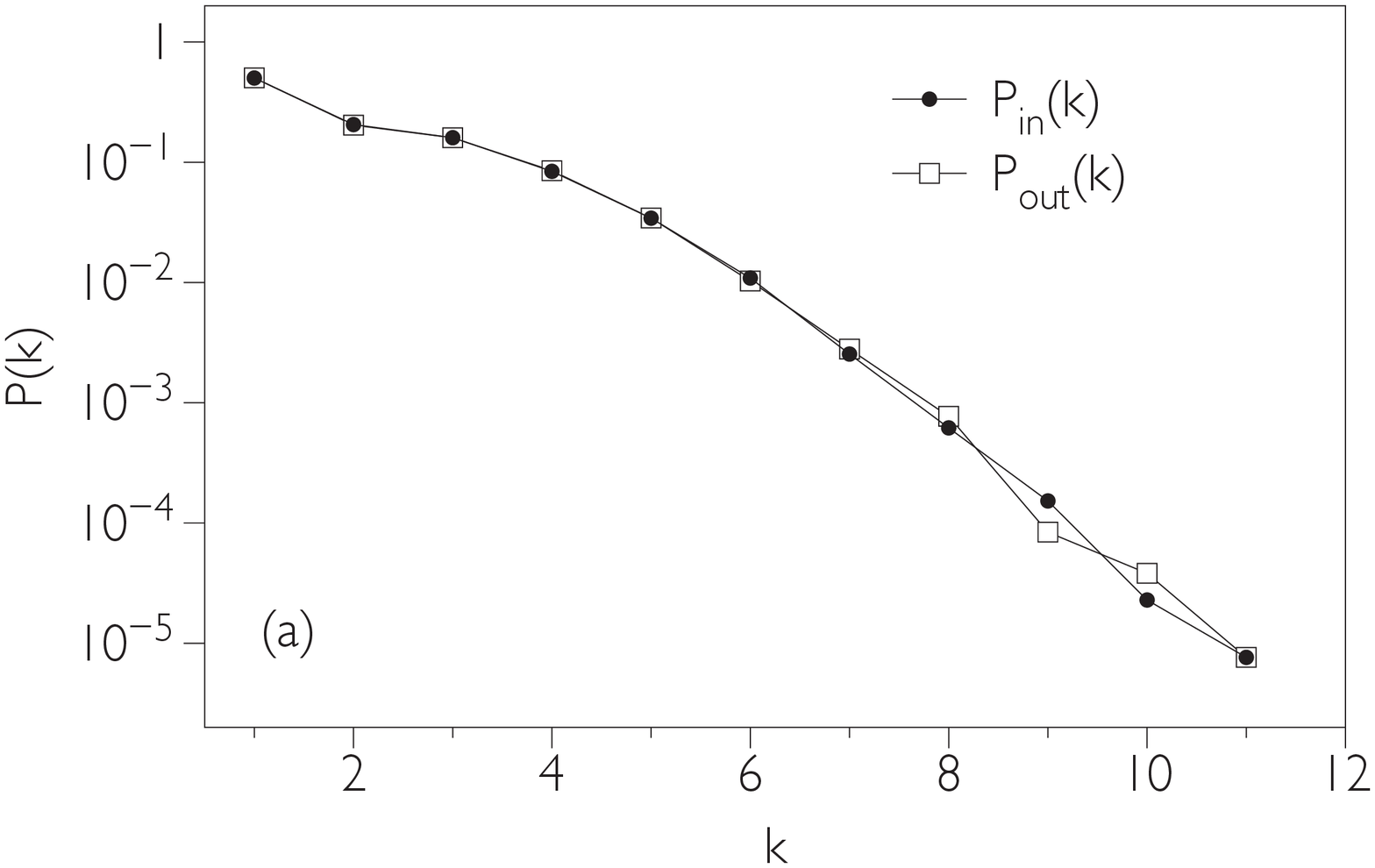}
\includegraphics[width=0.48\columnwidth]{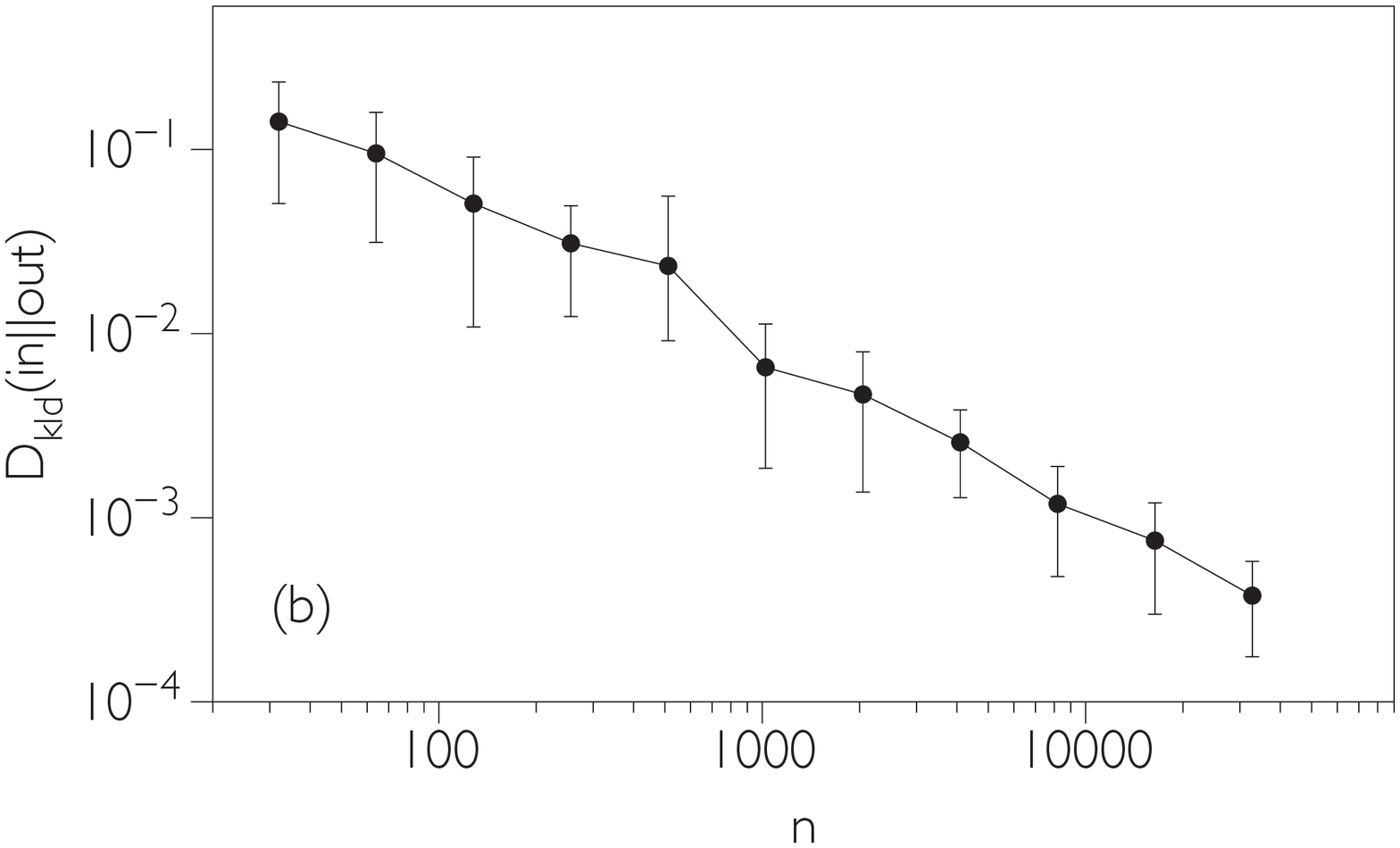}
\caption{{\bf Non-stationary unbiased (memoryless) additive random walk - HVG.}(a) Semi-log plot of the in and out degree distributions of the horizontal visibility graph associated to an unbiased random walk of $2^{17}$ steps generated from $x(t+1)=x(t)+\xi$, where $\xi \sim U[-0.5,0.5]$. Both distributions are identical up to finite-size effects fluctuations, suggesting that the underlying process is HVG-reversible. 
(b) Log-log plot of the irreversibility measure $D_{kld}(in||out)$ as a function of the series size $n$ (each dot is an average over 10 realizations). This measure vanishes asymptotically as $n^{-1}$, certifying that, albeit being a non-stationary process, it is HVG-reversible.}
\label{RW_HVG}
\end{figure}

In figure \ref{RW} and \ref{RW_HVG}, we plot the results of numerical simulations on the VG and HVG respectively, for an unbiased random walk case where $\xi\sim U[-0.5,0.5]$. We find that the {\it in} and {\it out} degree distributions of both graphs coincide (up to finite size effects), in good agreement with previous theorem. The irreversibility measure (panel b) for the HVG case decreases monotonically with series size $n$ as $O(1/n)$, yielding a vanishing value of irreversibility in the limit of large series. Roughly speaking, if we extended the relation between $D$ and entropy production to the non-stationary realm, we would conclude that the process described in equation \ref{RWmodel} has a null lower bound for its entropy production $dS/dt \geq D_{kld}(\text{in}||\text{out}) =0$, which is in good agreement with what is expected for a system which is in thermodynamic equilibrium.\\
\noindent The degree distributions for the VG have a power law decay $k^{-2}$, as reported in figure \ref{RW}. While we don't have a rigorous proof to support this, an heuristic derivation of this law can be outlined: the $k_{out}$ of a node $i$ (associated to $x_i$) chosen at random could be heuristically approximated as $P_{out}(k)\sim \#(k)q(k)$, where $q(k)$ defines the time window of the visibility basin (the average number of nodes that are 'visible' from $i$). As a rough approximation, $q(k)$ can therefore be related to the probability that the time series returns to $x_i$ after an excursion where $x<x_i$, and this is of order $k^{-3/2}$ for unbiased random walks (the first return distribution of an unbiased random walker). On the other hand, node $i$ won't necessarily have outgoing visibility will all and every node within the visibility basin, but just with a fraction of them. This fraction will depend on the fluctuations (roughness) of the time series within the basin. Roughness can be quantified in terms of the series standard deviation $\sigma$, which in unbiased random walks scale like $\sigma \sim t^{1/2}$. Accordingly, the percentage of $k$ nodes visible within the basin of visibility should be of order $k^{1/2}/k=k^{-1/2}$. Summing up, $P_{out}(k)\sim k^{-3/2}k^{-1/2}\sim k^{-2}$, in good agreement with the results found in the panel (a) of figure \ref{RW}.\\
Finally, note that the finite size fluctuations decrease in VG at a slower rate than for HVG, scaling with series size as $O(n^{-1/3})$. This is perhaps due to the fact that degree distributions in the VG case are power laws instead of exponential ones, thus finite size effects in this case case decay slower than for HVGs.

\subsection{Additive random walk with a drift}
\begin{figure}
\centering
\includegraphics[width=0.48\columnwidth]{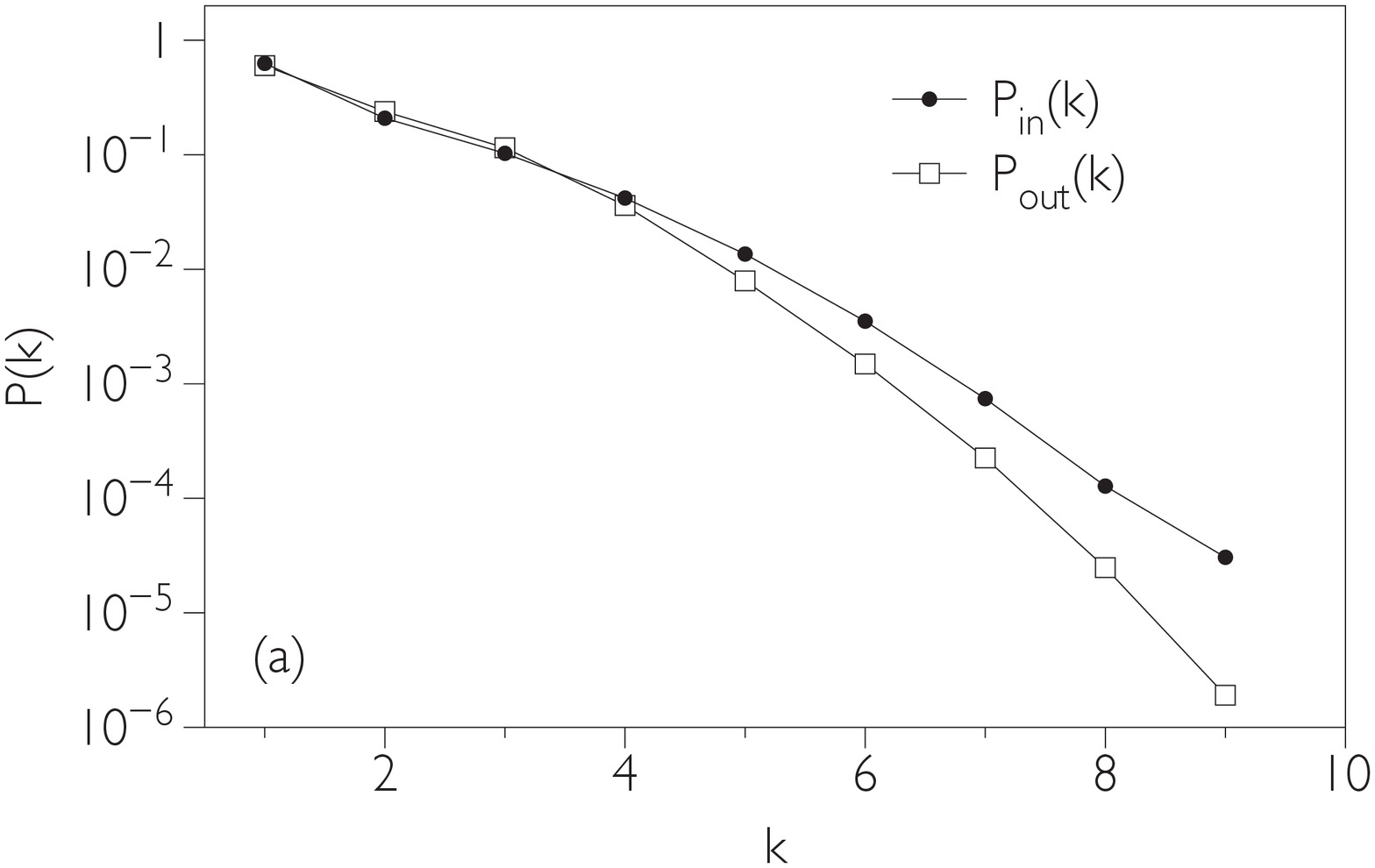}
\includegraphics[width=0.48\columnwidth]{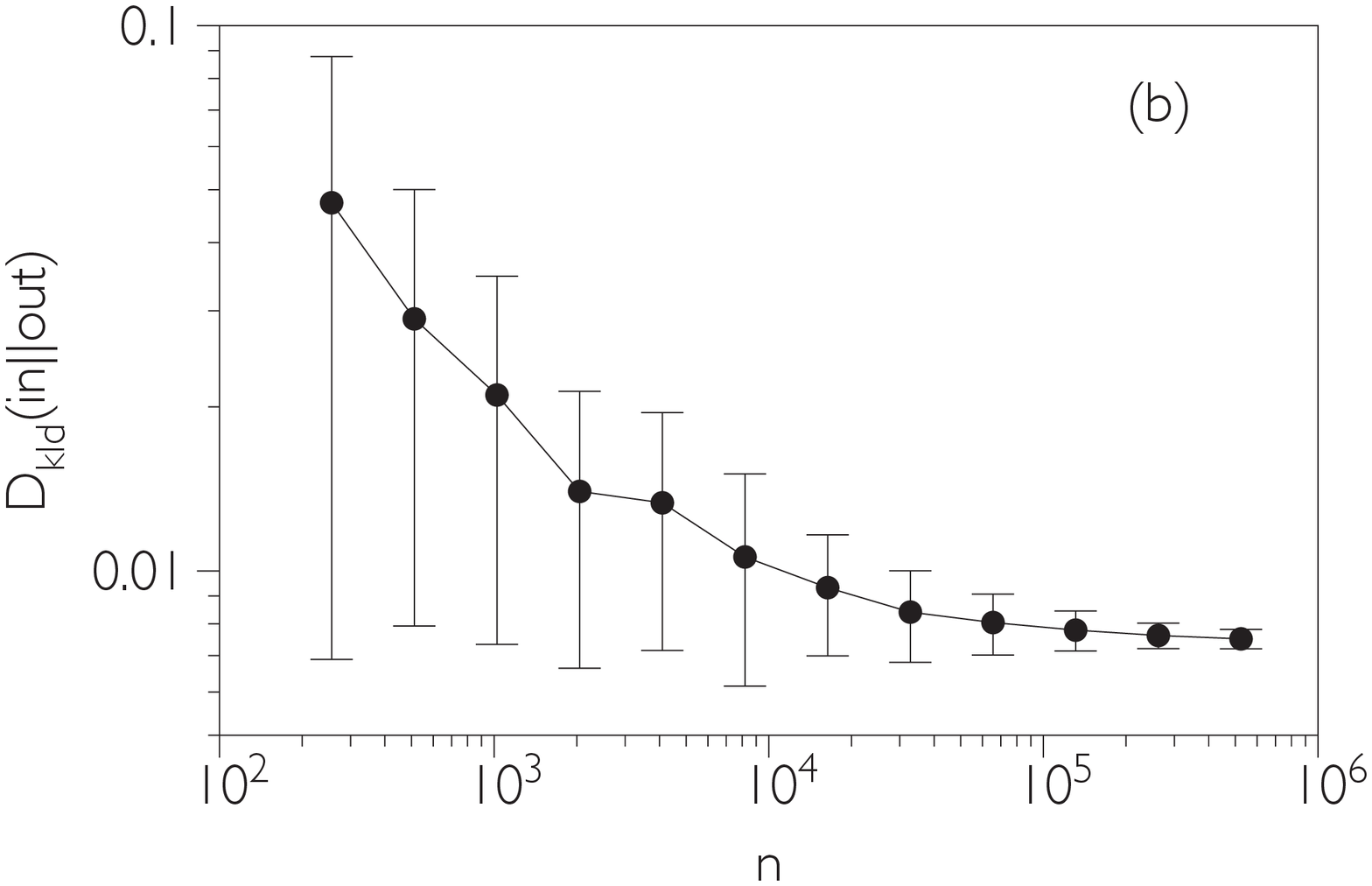}
\caption{{\bf Non-stationary (memoryless) additive random walk with drift - HVG.}(a) Semi-log plot of the in and out degree distributions of the horizontal visibility graph associated to an unbiased random walk of $2^{17}$ steps generated from $x(t+1)=x(t)+\xi$, where $\xi \sim U[-0.4,0.6],\ \langle \xi\rangle=0.1$. Distributions are different, suggesting that the process is HVG-irreversible.
(b) Log-log plot of the irreversibility measure $D_{kld}(in||out)$ as a function of the series size $n$ (each dot is an average over 10 realizations, and error bars denote the standard deviation). This measure converges asymptotically to a finite value, certifying that the process is HVG-irreversible.}
\label{RW_HVGdrift}
\end{figure}

In this subsection we explore the effect of adding a positive drift to an additive random walk. For that purpose, we bias equation \ref{RWmodel} by defining its increments as having a small positive mean:
$$x(t+1)=x(t)+\xi, \ \langle \xi\rangle>0$$
Note that this process is equivalent to superposing a linear trend, with positive slope $\langle \xi\rangle$ to the unbiased additive random walk described in equation  \ref{RWmodel}. Since the VG is invariant under addition of linear trends \cite{PNAS}, the VG associated to an unbiased random walk and a random walk with a linear trend is the same, so again this process VG-reversible (of course, by symmetry something similar happens in the case of a negative drift $\langle \xi\rangle<0$). Now, the HVG is not invariant under such transformation. Since the process is again VG and HVG stationary (choose $c=x_1-x_{1+\tau}$ in lemma 1), we should in principle be able to detect and quantify this additional source of irreversibility within the HVG setting. In figure \ref{RW_HVGdrift} we detail the numerical results for the HVG, for a concrete case where $\xi \sim [-0.4,0.6],\ \langle \xi\rangle=0.1$. The process is indeed HVG-irreversible. As the method provides a finite positive irreversibility value that converges to $\lim_{n\to \infty} D\approx 7.5 \cdot 10^{-3}$, time asymmetry for this non-stationary process can be quantitatively distinguished from the unbiased case, for both finite and infinite size series. Extending again the analogy between irreversibility and entropy production to the non-stationary realm, the HVG method would provide in this case a tighter bound $dS/dt\geq D_{kld}(in||out)\approx 7.5 \cdot 10^{-3}$.

\subsection{Non-Markovian additive random walk.} 
\begin{figure}
\centering
\includegraphics[width=0.48\columnwidth]{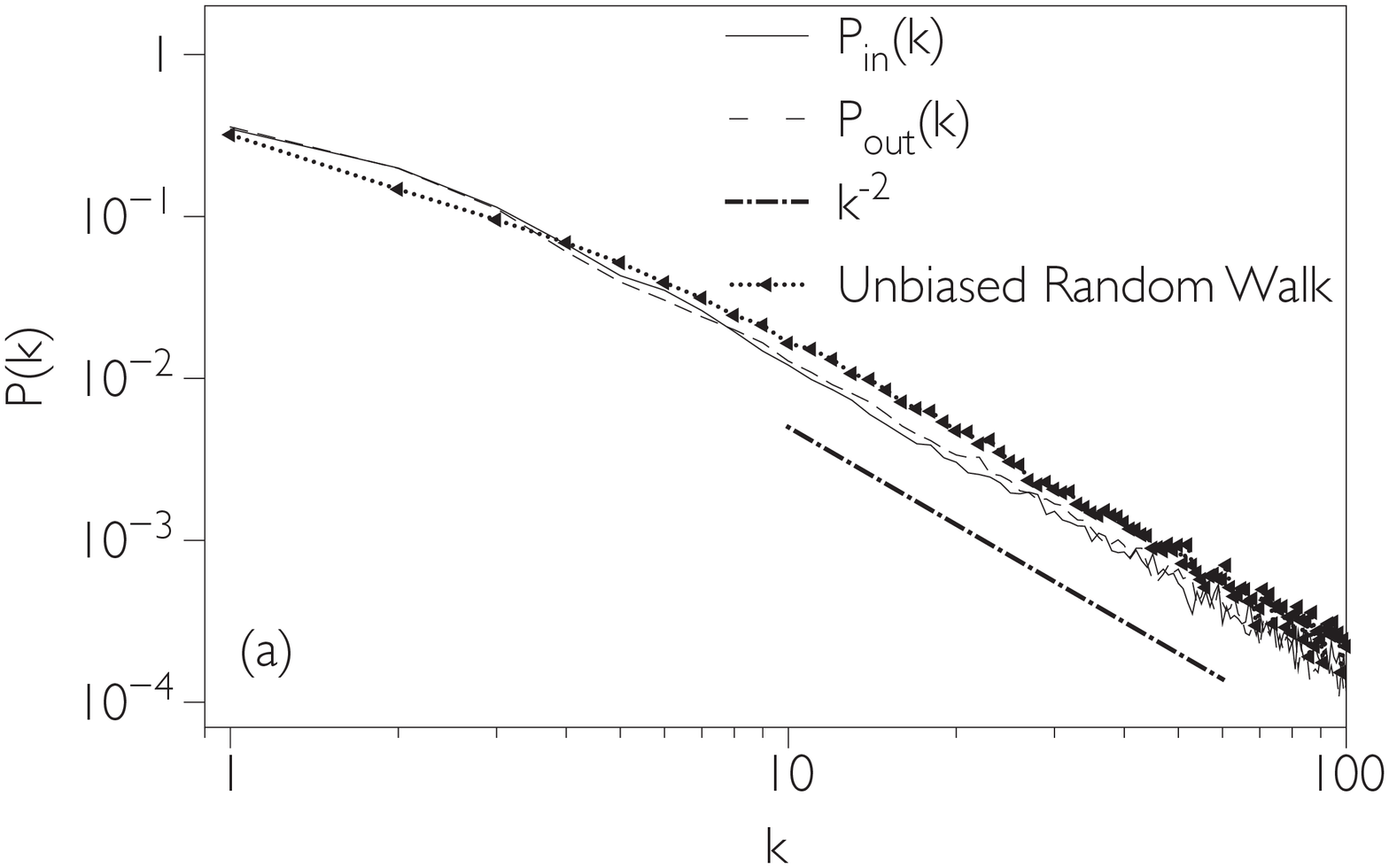}
\includegraphics[width=0.48\columnwidth]{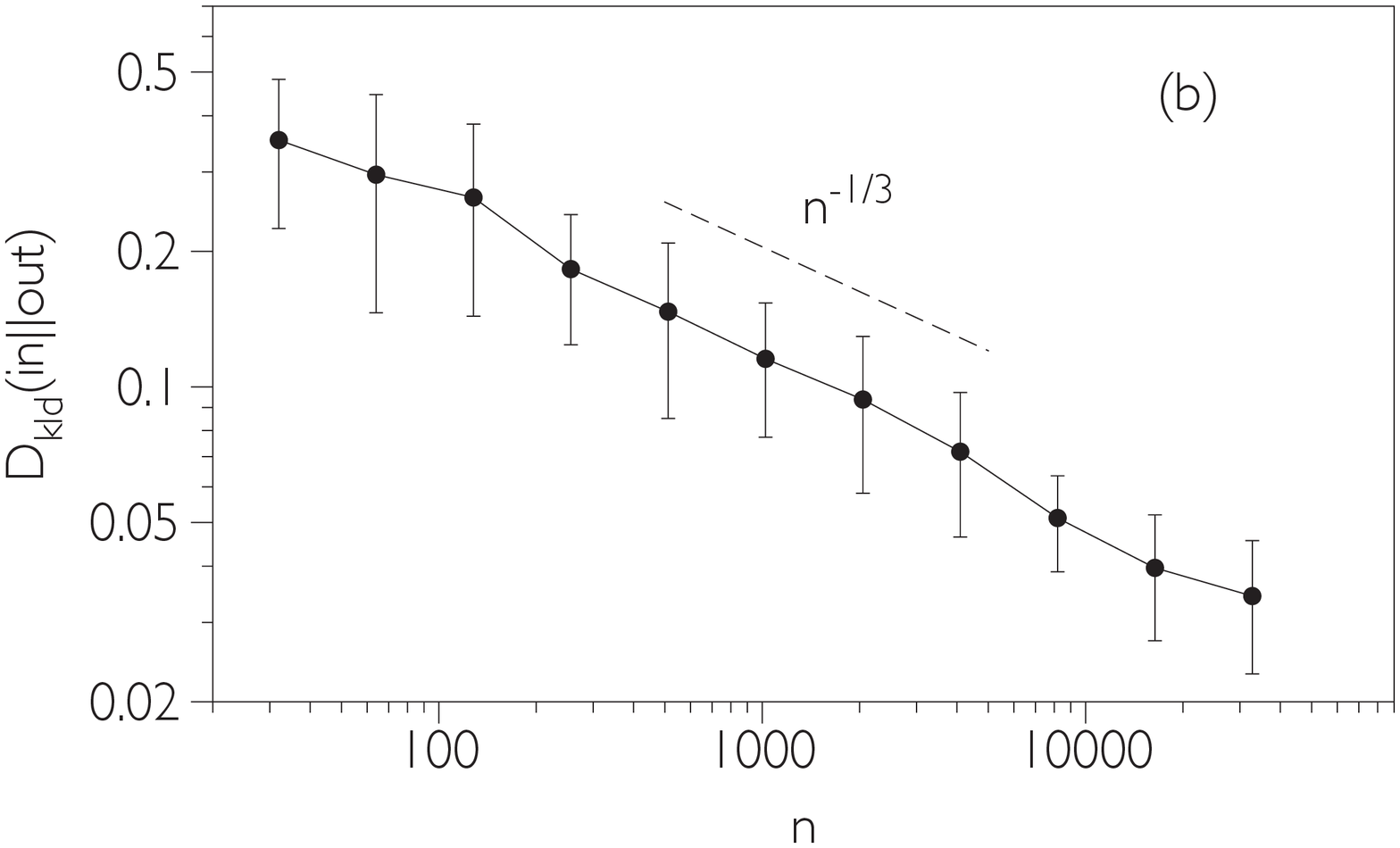}
\caption{{\bf Non-markovian additive random walk with memory - VG. }(a) Semi-log plot of the in and out degree distributions of the natural visibility graph associated to a biased random walk (see the text) of $2^{17}$ steps with delay $\tau=6$ and reset rate $r=0.3$. Both distributions are similar, suggesting that the onset of memory effects are not effectively captured by VG-reversibility, and although these are slightly different than for the baseline random walk, no major qualitative differences are observed. (b) Log-log plot of the irreversibility measure $D_{kld}(in||out)$ as a function of the series size $n$ (each dot is an average over 10 realizations). This measure vanishes asymptotically with series size as slowly as $n^{-1/3}$, so finite-size values can still be used for comparison with other models.}
\label{RWbiased}
\end{figure}

\begin{figure}
\centering
\includegraphics[width=0.48\columnwidth]{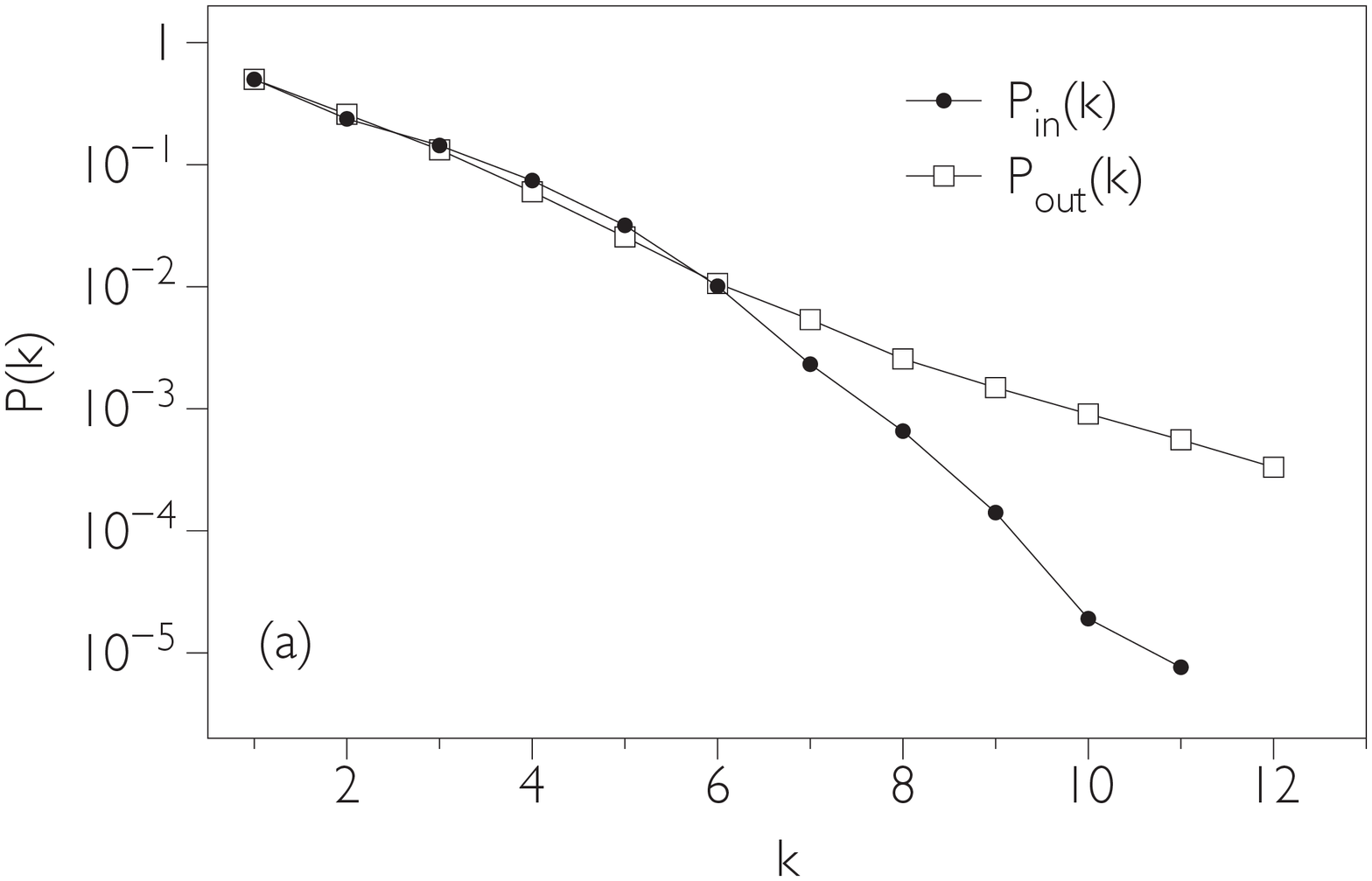}
\includegraphics[width=0.48\columnwidth]{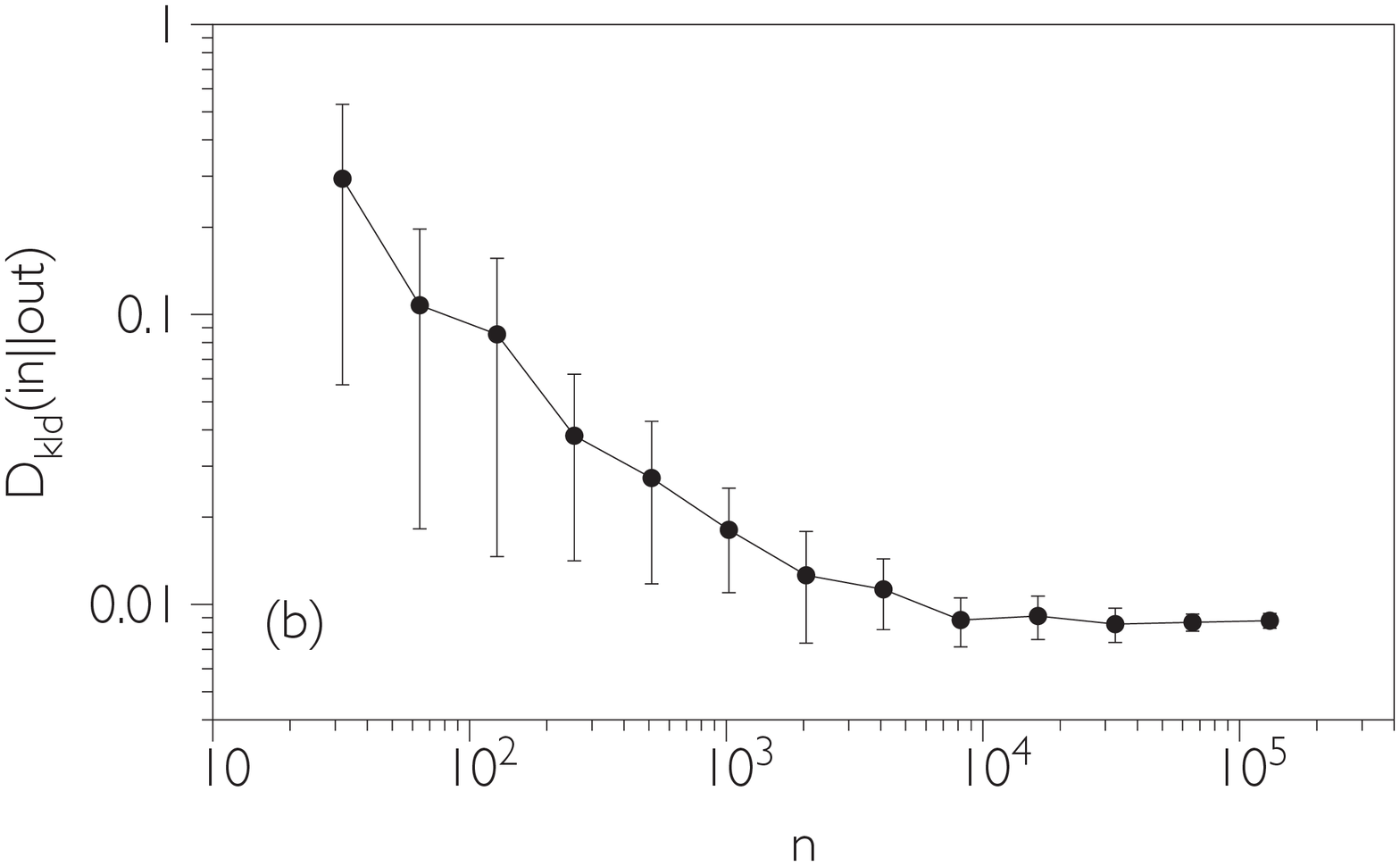}
\caption{{\bf Non-Markovian additive random walk with memory - HVG. }(a) Semi-log plot of the in and out degree distributions of the horizontal visibility graph associated to a biased random walk (see the text) of $2^{18}$ steps with delay $\tau=6$ and reset rate $r=0.3$. Both distributions are clearly different, suggesting that the onset of memory effects are effectively captured by the HVG. This seems to be a unique property of HVG (as VG fails to accurately capture this trait).
(b) Irreversibility measure $D_{kld}(in||out)$ as a function of the series size $n$ (each dot is an average over 10 realizations). This measure converges with increasing series size to a finite, non-null value, certifying that the process is HVG-irreversible. As the unbiased (memoryless) random walk (eq. \ref{RWmodel}) is in turn HVG-reversible, we conclude that the source of irreversibility captured in this process is only due to memory effects, as non-stationarities are filtered out.}
\label{RWbiasedHVG}
\end{figure}

Finally, let us consider the following generalization of a random walk:
\begin{equation}
x_{t+1} =
\left\{
\begin{array}{rcl}
     x_t + \xi & \textrm{if} & p > r
  \\ x_{t-\tau} & \textrm{if} & p<r
\end{array}
\right.
\label{heaviside}
\end{equation}
where for concreteness we set $\xi \sim U[-0.5,0.5]$, $r\in [0,1]$ is a fixed parameter that describes the reset rate, and $\tau \in \mathbb{N}$ is a fixed integer that describes the jump of the walker to previous states. This process can be used as a model for animal search in a 1d environment that includes memory of past locations \cite{denis}. It reduces to an unbiased Markovian random walk for $r=0$, and is non-Markovian for $r>0$ \cite{denis}.
In order to investigate the capacity of these methods to capture irreversibility associated to off-equilibrium dynamics, we have computed the in and out distributions of both VG and HVG associated to the non-Markovian random walk described in equation \ref{heaviside},
for a specific time delay $\tau=6$ and a resetting rate $r=0.3$ (note that other values can be chosen as well). Results are shown in figures \ref{RWbiased} and \ref{RWbiasedHVG}. The system is clearly HVG-irreversible. As the unbiased case is HVG-reversible, the mechanism responsible of triggering the irreversible character is not the process non-stationarity, but the onset of memory effects that drive the system away from equilibrium, 'producing entropy' at a rate $dS/dt \geq D_{kld}(in||out)\approx 8.8\cdot 10^{-3}$.\\
On the other hand, again the system appears to be VG-reversible again, failing to capture the source of irreversibility associated to the memory effects for asymptotic large sizes. However, note that the convergence speed of this process is rather slow (with $D\sim n^{-0.35}$ as reported in figure \ref{RWbiased}), what permits us to compare finite-size irreversibility values. For instance, for $n=10^3$, $D\sim O(10^{-1})$, to be compared with the much smaller analogous result for unbiased random walk $D\sim O(10^{-2})$. We can conclude that, although the VG fails asymptotically to detect irreversibility in this process, finite-size values can still be used in practice to compare the degree of reversibility with other processes.\\

\section{Multiplicative random walks}
We now explore the properties of VG and HVG associated to certain random multiplicative processes. A multiplicative random walk (MRW) is a stochastic process $x(t)$ that follows the equation
\begin{equation}
x(t+1)=\xi\cdot x(t),
\label{MulRW}
\end{equation}
where $\xi$ is a random variable extracted from some distribution. If we identify ${\cal X}\equiv\log x$ and $\eta\equiv\log \xi$, this process is formally equivalent to an additive random walk in logarithmic space, as equation \ref{MulRW} reduces to
$${\cal X}(t+1)={\cal X}(t)+ \eta.$$
A standard assumption is that ${\cal X}$ approaches a lognormal distribution accordingly. However, we should be very cautious at this point, as the properties of an additive random walker provided by the central limit theorem are not directly applicable to the MRW in logarithmic space, due to non-ergodicity and to the relevant effect of extreme events \cite{redner}, which preclude self-averaging and convergence to the asymptotic lognormal distribution. Also, note that ${\cal X}$ only exists as the logarithm of $x$ for $x>0$, what in turn imposes restrictions on the support of $\xi$.\\

\begin{figure}
\centering
\includegraphics[width=0.6\columnwidth]{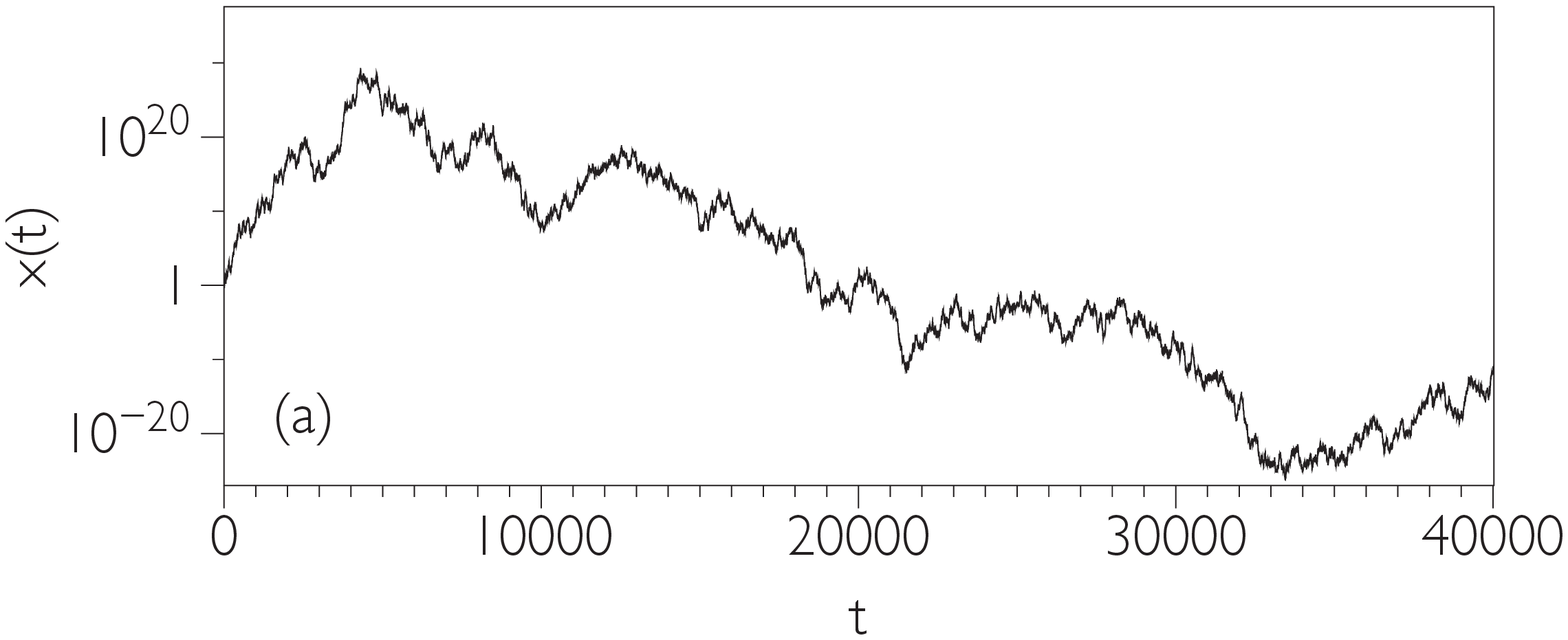}
\includegraphics[width=0.58\columnwidth]{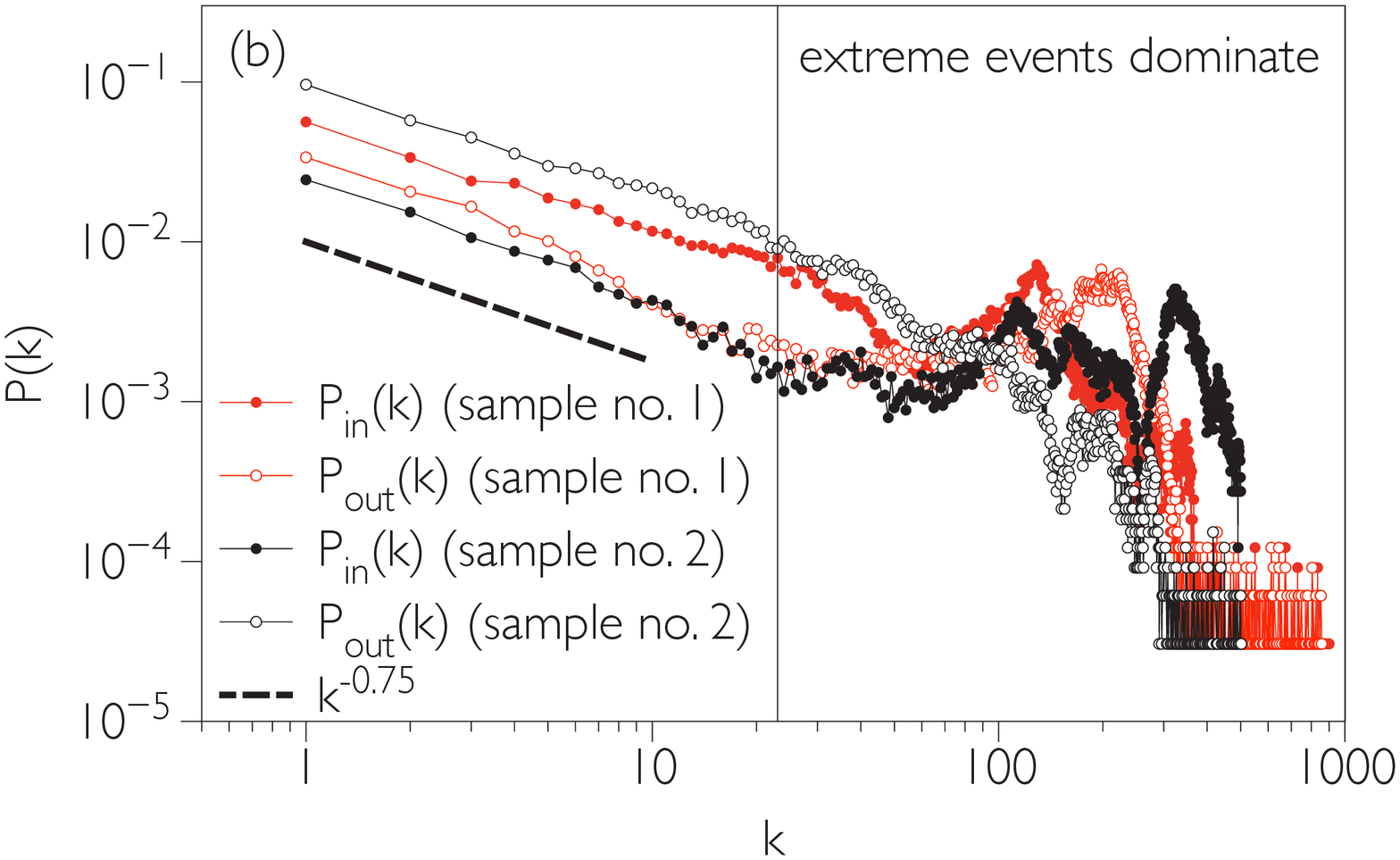}
\caption{(a) Log-linear plot of a sample time series generated through the process $x(t+1)=\xi\cdot x(t), \ \log \xi \sim U[-0.5,0.5]$. This multiplicative process is additive in logarithmic space, hence in a log-linear plot, the time series looks similar to an additive random walk with no drift, as $\log \xi$ is uniformly distributed in a symmetric interval so $\langle \log\xi\rangle=0$. Irreversibility measures on $x(t)$ are depicted in figure \ref{MRW}. (b) (Color online) Log-log plot of the {\it in} and {\it out} degree distributions for two different realizations of $2^{15}$ data of a multiplicative random walk $x(t+1)=\xi x(t)$, where $\log \xi \sim U[-0.5,0.5]$. The curves have a power law decay with a fairly stable exponent $k^{-0.75}$, followed by wildly fluctuating tails. These are related to the presence of extreme events in the series, which are exponentially rare but exponentially large, and dominate the tails \cite{redner}.
}
\label{MRWseries}
\end{figure}

\noindent \underline{\bf Case 1. \\}
\noindent Here we explore two simple versions of a MRW. In the first case we set $\log \xi$ to be uniformly distributed in $[-0.5,0.5]$, so $\xi \sim \exp(s-0.5), \ s\sim U[0,1]$. With positive initial condition for $x$, each realization of this MRW is thus qualitatively similar to a realization of an additive unbiased uniform random walk in logarithmic space (see figure \ref{MRWseries} for an illustration). Notice that in the figure $x(t)$ wildly varies on amplitude, reaching values in the interval $[10^{-20},10^{20}]$ for $n=4\cdot 10^4$, however the sketch is, in log-linear scale, qualitatively similar to a realization of an additive unbiased random walk.\\
At this point we need to introduce and prove an additional property of HVGs:\\

\noindent {\bf Definition. } A VG (HVG) is invariant under monotonic transformations if the VG (HVG) graphs associated to a time series $\{x_1,x_2,\dots,x_n\}$ and $\{f(x_1),f(x_2),\dots,f(x_n)\}$ (where $f(x)$ is an arbitrary monotonic function) are identical.\\

\noindent {\bf Proposition 1.} VGs are in general not invariant under monotonic transformations.\\
{\it Proof: } The rationale is that the visibility link criterion is a convexity one, so any monotonic transformation that alters the convexity properties of the series will alter the resulting VG. We give here two counterexamples. Consider the time series ${\cal S}_1=\{1,2\dots,n\}$ and $f(x)=x^3$, such that ${\cal S}_2=\{f(1),f(2),\dots,f(n)\}$. The VG associated to ${\cal S}_1$ is a chain graph (1d lattice) whereas the VG associated to ${\cal S}_2$ is a complete graph. Consider now an additional series ${\cal S}_3=\{ \log f(1), \log f(2),\dots,\log f(n)\}=3\{\log 1, \log 2, \dots, \log n\}$. Trivially, the VG associated to the concave series ${\cal S}_3$ is again a chain graph , which is different from the VG of ${\cal S}_2$.  $\square$\\
  
\noindent {\bf Proposition 2.} HVG are invariant under monotonic transformations.\\
{\it Proof:} The link criterion for HVGs is solely based on the specific {\it ordering} of the data, not on their values. Hence HVG shall be are invariant under order-preserving transformations.  Monotonic functions are indeed isotone mappings, so order-preserving. Consider the time series ${\cal S}_1=\{x_1,x_2,\dots,x_n\}$ and ${\cal S}_2=\{f(x_1),f(x_2),\dots,f(x_n)\}$. The link criterion for HVG in ${\cal S}_1$ is:\\
$x_k<\inf(x_i,x_j),\ \forall k: i<k<j$. But if this criterion is fulfilled, then we have
$f(x_k)<\inf(f(x_i),f(x_j)),\ \forall k: i<k<j$ if $f$ is monotonic.  Therefore two connected nodes $i$ and $j$ in the HVG associated to ${\cal S}_1$ yield two connected nodes in the HVG associated to ${\cal S}_2$, which make both HVGs identical. $\square$\\

\begin{figure}
\centering
\includegraphics[width=0.48\columnwidth]{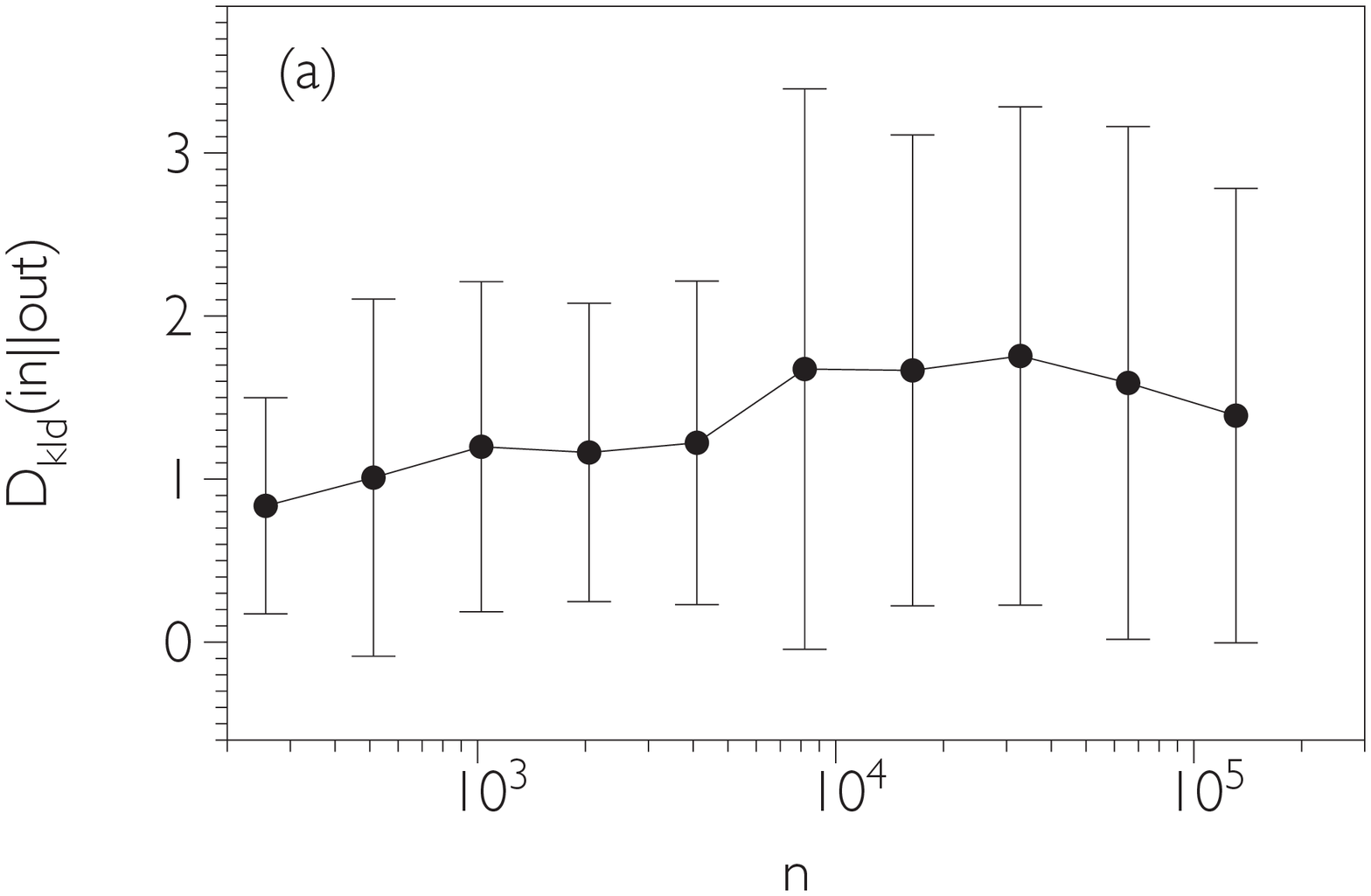}
\includegraphics[width=0.48\columnwidth]{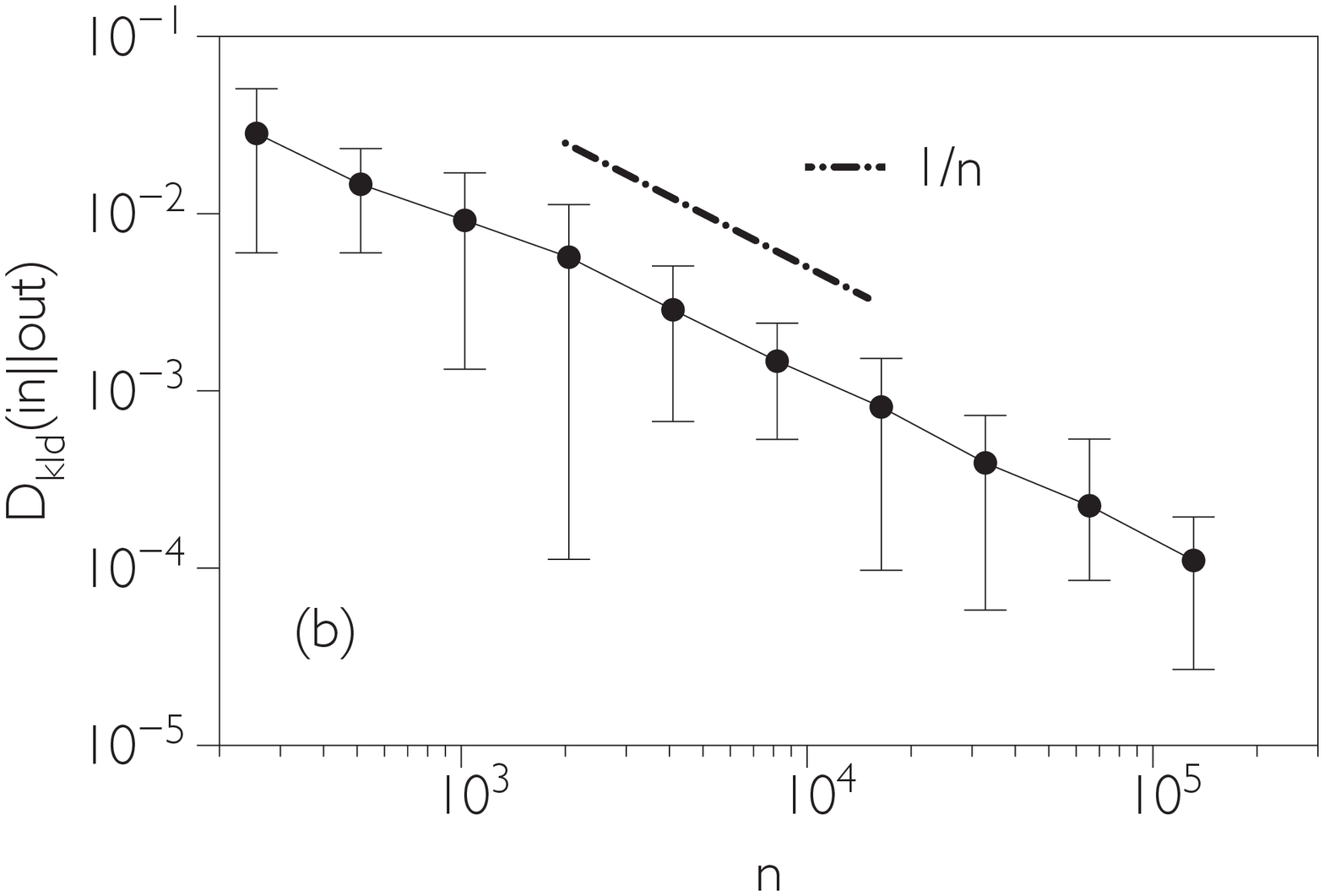}
\caption{{\bf Multiplicative random walk with uniformly distributed log-returns. }(a) Linear-log plot of the irreversibility measure $D_{kld}(in||out)$ as a function of the series size $n$ (each dot is an average over 100 realizations and error bars account for $\pm \sigma$) computed from the VG associated to a multiplicative random walk $x(t+1)=\xi \cdot x(t), \ \log\xi \sim U[-0.5,0.5]$. The measure converges to a finite value, so the process is VG-irreversible. (b) Log-log plot of the same measure computed from the HVG. The measure decays with series size $n$, so the process is HVG-reversible.}
\label{MRW}
\end{figure}

\noindent In the light of the previous propositions, for $f(x)=\log x$, one finds that the HVG is equivalent to the HVG associated to a realization of an unbiased additive random walk, i.e. the process is HVG-stationary and HVG-reversible. This is confirmed by the vanishing values of $D$ in panel (b) of figure \ref{MRW} (ensemble averaged over 100 realizations). On the other hand, by proposition 1, VG is not in general invariant under log-transformations, so the VG does not reduce here to the one found in the additive case.  In panel (a) of figure \ref{MRW} we plot the numerical results of $D$ computed from the VG. First, $D$ does not vanish with system's size, suggesting VG-irreversibility. Interestingly enough, values highly fluctuate within each ensemble average, as denoted by large standard deviations. This lack of self-averaging is related to the non-ergodic nature of the MRW: in panel (b) of figure \ref{MRWseries} we plot, in log-log scales, the {\it in} and {\it out} degree distributions of two different realizations of the process, for $n=2^{15}$. The shape of the distributions all begin with a power law decay $k^{-0.75}$, followed by a tail that evidence large fluctuations. This is indeed the part of the distributions that vary from realization to realization, and as it is ruled by extreme events (exponentially rare but exponentially different), deviations from the stationary distribution are large. This effect is well-known in multiplicative random processes \cite{redner}, and precludes us to interchange average values with most probable ones (this is also the main reason why convergence of this process to the lognormal distribution is not straightforward). Because of that, distributions do not converge smoothly for large series to their asymptotic form, finding large standard deviations in the estimation of $D$.\\

\begin{figure}
\centering
\includegraphics[width=0.6\columnwidth]{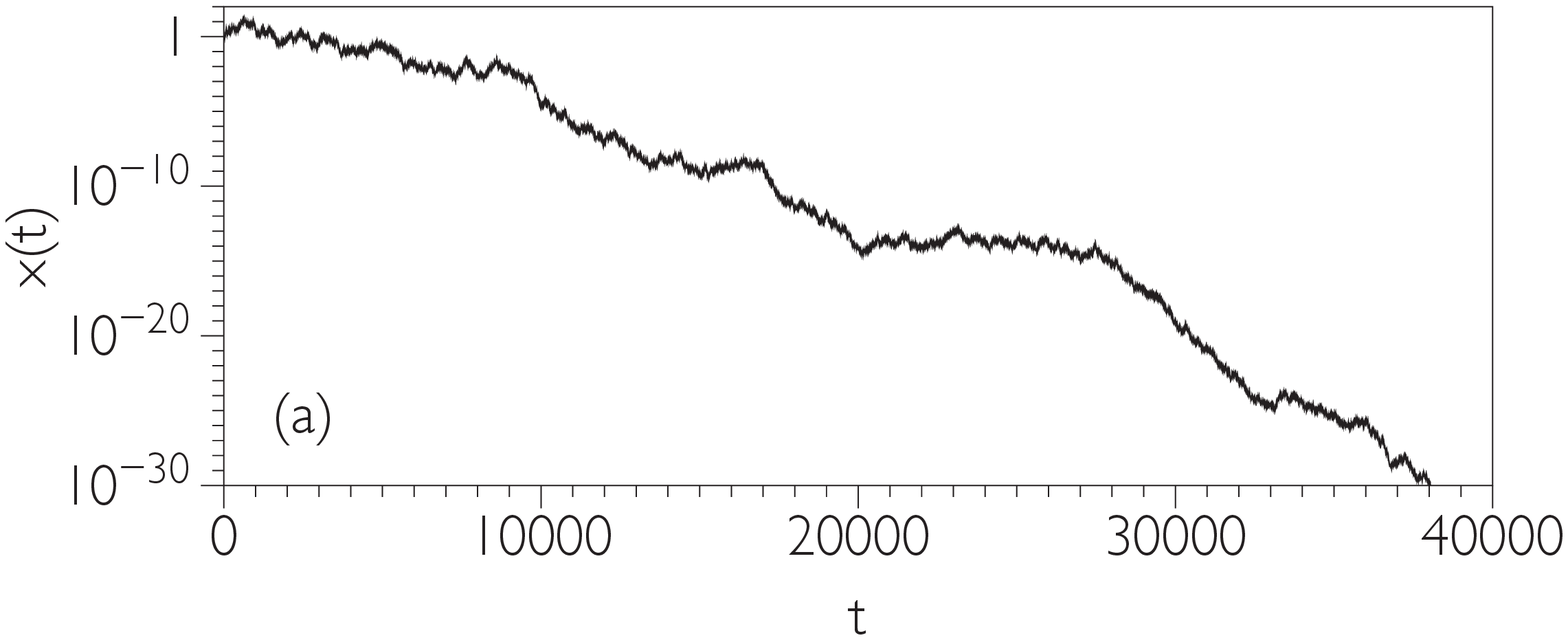}
\includegraphics[width=0.58\columnwidth]{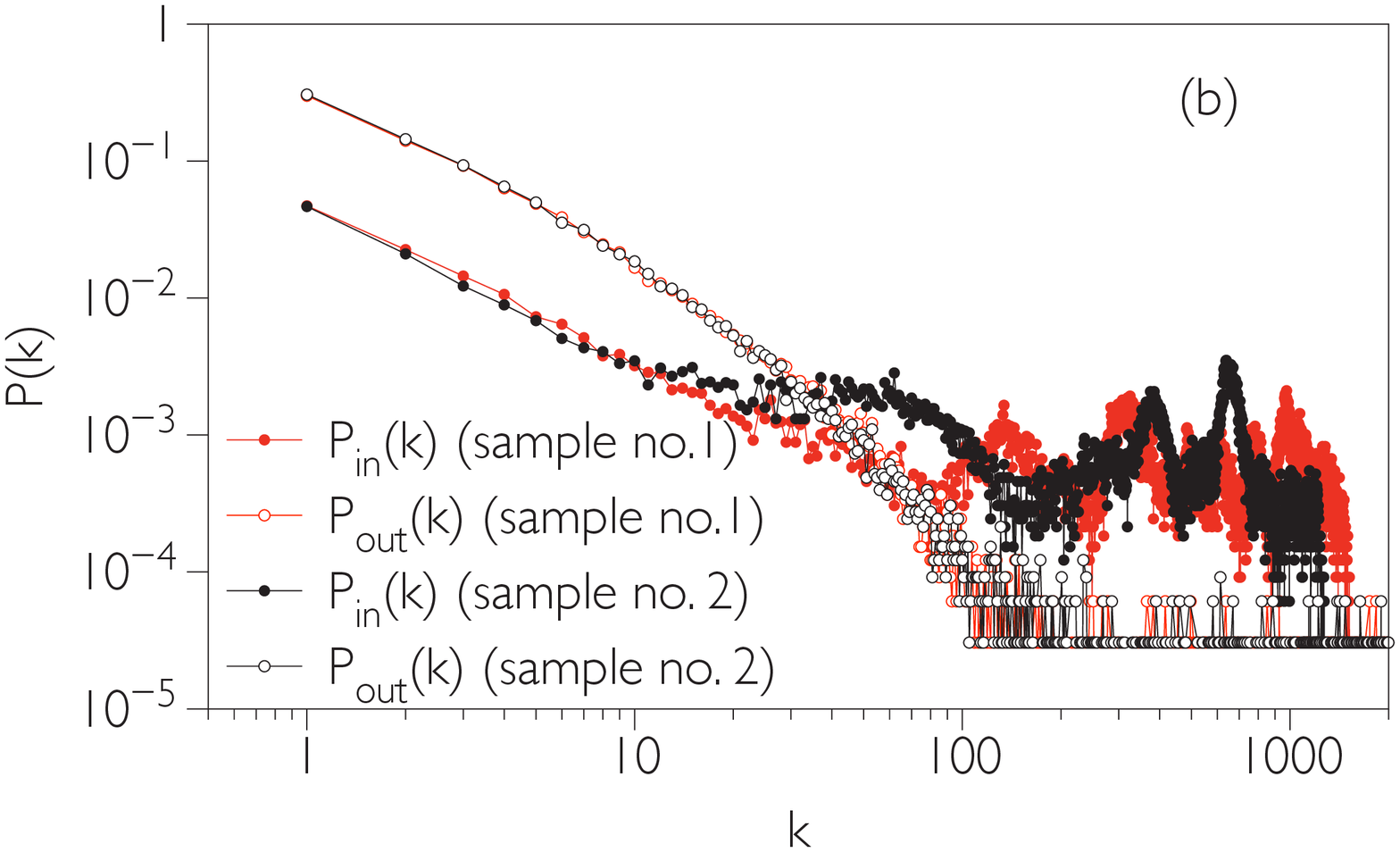}
\caption{(a) Log-linear plot of a sample time series generated through the process $x(t+1) = \xi \cdot x(t)$, $\xi \sim U[0.9,1.1]$, so in logarithmic space the process shares similarities with an additive random walk with a negative drift, as $\langle \log\xi\rangle<0$. Irreversibility measures on $x(t)$ are depicted in figure \ref{MRWdrift}.
(b) (Color online) Log-log plot of the {\it in} and {\it out} degree distributions for two different realizations of $2^{15}$ data of a multiplicative random walk $x(t+1)=\xi x(t)$, where $\xi \sim U[0.9,1.1]$.}
\label{MRWseries2}
\end{figure}

\begin{figure}
\centering
\includegraphics[width=0.48\columnwidth]{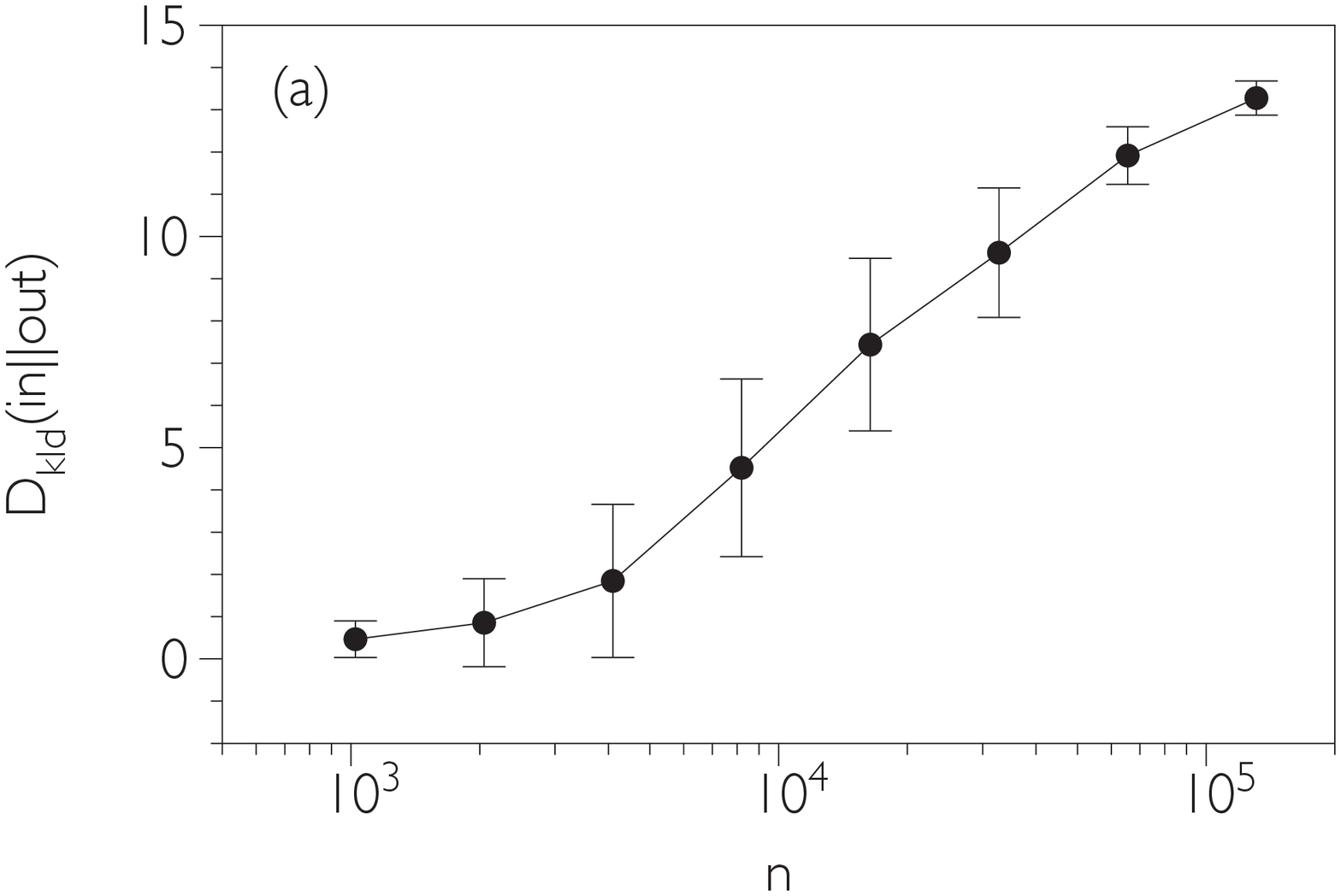}
\includegraphics[width=0.48\columnwidth]{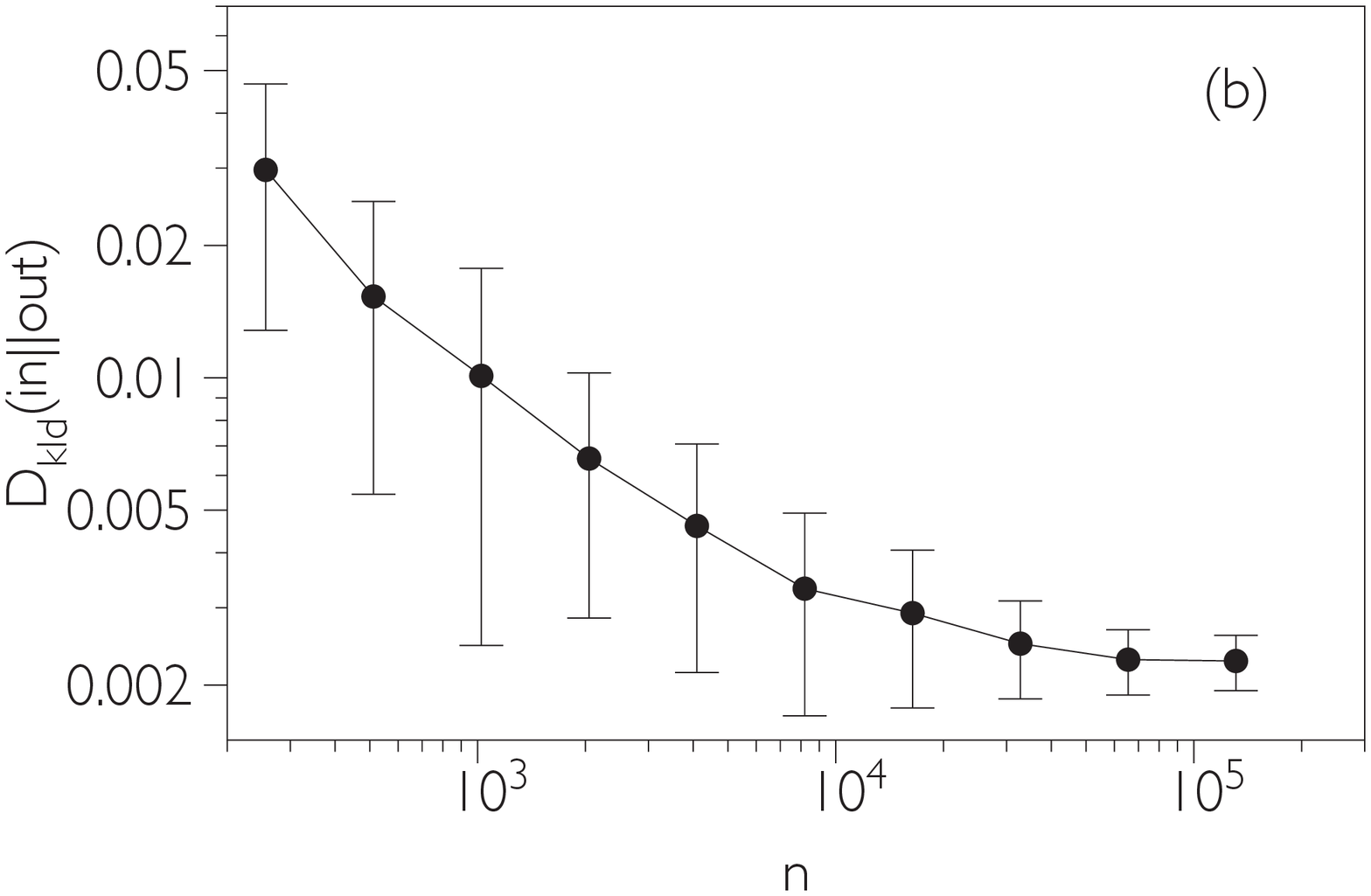}
\caption{{\bf Multiplicative random walk with drift. }(a) Log-linear plot of the irreversibility measure $D_{kld}(in||out)$ as a function of the series size $n$ (each dot is an average over 100 realizations and error bars account for $\pm \sigma$) computed from the VG associated to a multiplicative random walk $x(t+1)=\xi \cdot x(t), \ \xi \in U[0.9,1.1]$, which induces an additive random walk with drift in logarithmic space. The measure diverges logarithmically with series size, hence the process is infinitely VG-irreversible. (b) Log-log plot of the same measure computed from the HVG. The measure converges to a finite, non-null value with series size $n$, hence the process is HVG-irreversible.}
\label{MRWdrift}
\end{figure}

\noindent \underline{\bf Case 2. \\}
In the second case, as an example of a MRW with symmetrical multiplicative noise, we set $\xi$ to be uniformly distributed in $U[0.9,1.1]$. With positive initial condition for $x$, each realization of this MRW is qualitatively similar to an additive random walk with a small negative drift in logarithmic space (see figure \ref{MRWseries2} for an illustration). The reason is that $\eta \sim \log s, s\sim U[0.9,1.1]$, thus $\langle \eta \rangle <0$, what implies that $x(t)\to 0$ for large values of $t$. This is a clear-cut case of an irreversible process.  By virtue of proposition 2, the HVG should now be qualitatively similar to the additive random walk with drift case, hence HVG-irreversible. Indeed, we find that the process is finitely HVG-irreversible (panel (b) of figure \ref{MRWdrift}). On the other hand, the analysis based on VG is again not reduceable (by proposition 1) to the additive case. In order to explore the properties of VG, we need to advance yet another property:\\

\begin{figure}
\centering
\includegraphics[width=0.6\columnwidth]{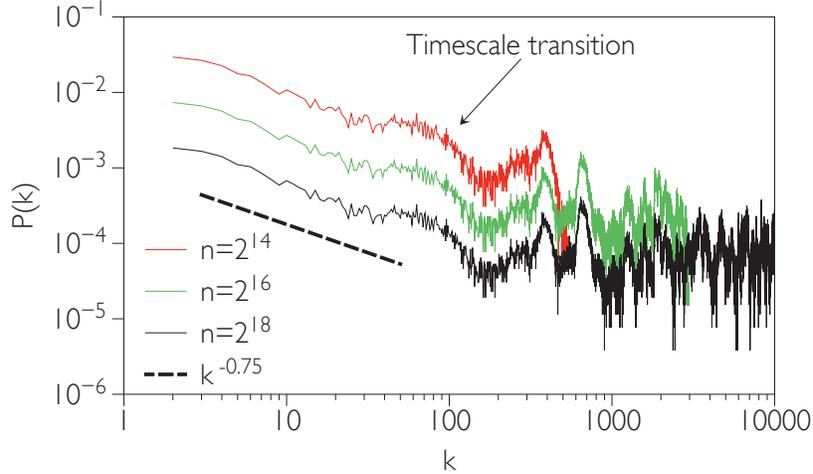}
\caption{(Color online) Log-log plot of the degree distribution of a VG associated to the first $2^{n}$ data (where, from up to bottom, $n=14,16,18$) of a time series generated via the MRW with symmetric multiplicative noise $\xi \sim U[0.9,1.1]$, which is qualitatively similar to an additive random walk with negative drift in logarithmic space. There is a power-law contribution at small degrees (associated to the fast timescale) and a steady wave-fluctuating part associated to the envelope whose extension increases with series size (see the text for details).}
\label{scales}
\end{figure}

\noindent \noindent {\bf Proposition 3.} Let the time series ${\cal S}=\{x_1,x_2,\dots,x_n\}$ be such that there exist a convex function $f$ such that $x_i=f(i) \ \forall x_i$. Then the VG associated to ${\cal S}$ is the complete graph $K(n)$. If instead of convex, $f$ is concave, then VG is the chain graph (1d lattice).\\
{\it Proof:} Time series generated by convex functions generate complete visibility graphs. This can be easily proved geometrically. Consider two arbitrary data $x_i,x_j \in {\cal S}$ where without loss of generality $i<j$, and consider the segment that links $x_i$ and $x_j$. As $f$ is convex, the slope of this segment is always larger 
than the slope of any segment connecting $x_i$ and $x_k$ for $i<k<j$:
\begin{eqnarray}
\frac{x_j-x_i}{j-i}>\frac{x_k-x_i}{k-i}
\label{slope}
\end{eqnarray}
Equation \ref{slope} is indeed equivalent to the visibility link criterion, so $i$ and $j$ are connected. Since this holds $\forall i,j \in [1,n]$, then the resulting visibility graph is the complete graph $K(n)$. The proof for the concave function follows analogously. $\square$\\

\noindent Of course, a time series extracted from the MRW with symmetrical noise cannot be simply represented as the graph of a convex function - the MRW is not differentiable to begin with, and although there is a large-scale negative trend, there are also episodes of uphill fluctuations. However, as this process is akin to an additive random walk with a negative drift in logarithmic space, then, roughly speaking, the {\it envelope} of $x(t)$ can be approximated by $t^{-\delta}$, which is convex. Thus as a very crude approximation, we should expect that the degree distribution of VG is the result of two competing time scales: at slow timescales there is bulk of very largely connected nodes (this is the envelope), whereas the time series random fast fluctuations contribute with a random-walk like part $k^{-0.75}$. The envelope is not strictly convex, but it displays a composition of a convex function and a wave-like structure (associated to small scale uphill fluctuations). These uphill fluctuations work as effective visibility barriers, hence we should expect that the tail of the distribution is not delta-like (the result of a convex envelope) but a somewhat wavy distribution. As series size increases, a larger amount of data contribute to the envelope effect, hence we expect an increase of the slow timescale region and, by normalization, vertical shift in the distribution. We have run numerical simulations to explore this behavior. We have generated a time series of $2^{18}$ data and computed the VG associated to time windows of the first $n=2^{14}, 2^{16}$ and $2^{18}$ data. In figure \ref{scales} we plot, in log-log scales, the degree distribution of this VG for different series size. The two contributions as well as the transition between the two timescales can be clearly seen.
(Note that for other realizations the shape associated to slow timescales will largely vary from what reported in the figure, although the effect of series increase would be similar).\\

\noindent Getting back to the {\it directed} VG, notice that there is a large asymmetry between $k_{in}$ and $k_{out}$, in direct relation to the fact that the time series is decreasing. Heuristically, we expect that few nodes have very large $k_{out}$ (and these corresponding to early nodes - associated to early data) that have {\it out} visibility of a large part of the long, asymptotically vanishing series. On the other hand, a representatively large portion of the nodes will have large $k_{in}$ (nodes associated to smaller and smaller values that receive links from past nodes). Actually, this percentage will increase as the series size increases, and will largely fluctuate in consonance with the fluctuations of the time series - hence we expect large fluctuations from sample to sample. All these features are confirmed in panel (b) of figure \ref{MRWseries2}. On the other hand, the wild fluctuations at the level of the {\it in} degrees will generate finite irreversibility measures. Moreover, as the difference between {\it in} and {\it out} distributions are expected to increase with series size, the irreversibility measure should also be an increasing function of series size. This is confirmed in panel (a) of figure \ref{MRWdrift}) where the irreversibility measure increases with series size with no apparent bound, hence suggesting that the process is asymptotically infinitely VG-irreversible.\\



\section{Conclusions}
In this work we have investigated, via analytical calculations and numerical simulations, the properties of visibility and horizontal visibility graphs associated to several non-stationary stochastic processes, as well as their ability to quantify several degrees of time irreversibility. We have proved that unbiased additive random walks, while non-stationary, are both VG/HVG-stationary and VG/HVG-time reversible, reconciling the fact that Brownian particles in thermodynamic equilibrium do not produce entropy on average. On the other hand biased memoryless additive random walks are HVG-irreversible with finite irreversibility measures that quantify the degree of time asymmetry, while these are still VG-reversible, as VG is invariant under superposition of linear trends in the original data. Numerics suggest that HVG can capture for both finite and infinite series size, the irreversible nature of non-Markovian additive random walks, whereas VG is only able to do so for finite series. For multiplicative random walks, the processes are HVG-reversible if the process is akin to an unbiased additive process in logarithmic space, and time irreversible if the process reduces to a biased additive process in logarithmic space. These latter results hold as HVG is invariant under monotonic transformations. Finally, the VG capture the time irreversible character of multiplicative random walks, yielding finite values in the unbiased case and asymptotically diverging quantities in the biased case.\\
While most of these are conclusions based on the asymptotic behavior (i.e. in the limit of infinitely long time series), it should be noted that finite size time series always yield finite, non-null values of HVG and VG irreversibility. As the convergence speed for reversible processes is rather slow ($O(1/n)$ for HVGs and $O(n^{-0.4})$ for VGs), these finite-size values can still be used in practice to compare the degree of statistical HVG/VG irreversibility for finite samples. This fact enables the use of VG/HVG methods in empirical (hence finite) datasets.\\
As a final remark, these analysis suggest that the horizontal visibility method seems to be better suited to capture irreversibility traits in additive non-stationary signals (I(1) processes), whereas the visibility method might be a better tool to quantify these signals which are better modeled by multiplicative models. These results should be taken into account when assessing time irreversibility via visibility graphs, and will be useful particularly for the analysis of empirical non-stationary signals, such as financial, geophysical, or biological data.\\

{ \bf Acknowledgments.} We thank an anonymous referee for interesting comments.

\bibliography{apssamp}

\end{document}